 \documentclass[final,5p,times]{elsarticle}

\usepackage{amssymb}
\usepackage{xcolor}
\usepackage{graphicx}%
\usepackage{multirow}%
\usepackage{amsmath,amssymb,amsfonts}%
\usepackage{amsthm}%
\usepackage{mathrsfs}%
\usepackage[title]{appendix}%
\usepackage{xcolor}%
\usepackage{textcomp}%
\usepackage{manyfoot}%
\usepackage{booktabs}%
\usepackage{algorithm}%
\usepackage{algorithmicx}%
\usepackage{algpseudocode}%
\usepackage{listings}%
\usepackage{physics}
\usepackage{float}
\usepackage{subcaption}
\usepackage{stfloats}
\usepackage{makecell}

\journal{arXiv}

\newtheorem{remark}{Remark}

\newtheorem{proposition}{Proposition}

\begin{document}

\begin{frontmatter}



\title{Feedback-Based Quantum Strategies for Constrained Combinatorial Optimization Problems}


\author[first]{Salahuddin Abdul Rahman}
\affiliation[first]{organization={Aalborg University},
            city={Aalborg},
            country={Denmark}}
\author[Second]{Özkan Karabacak}
\affiliation[Second]{organization={Kadir Has University},
            city={Istanbul},
            country={Turkey}}

\author[first]{Rafal Wisniewski}

\begin{abstract}
Feedback-based quantum algorithms have recently emerged as potential methods for approximating the ground states of Hamiltonians. One such algorithm, the feedback-based algorithm for quantum optimization (FALQON), is specifically designed to solve quadratic unconstrained binary optimization problems. Its extension, the feedback-based algorithm for quantum optimization with constraints (FALQON-C), was introduced to handle constrained optimization problems with equality and inequality constraints. In this work, we extend the feedback-based quantum algorithms framework to address a broader class of constraints known as invalid configuration (IC) constraints, which explicitly prohibit specific configurations of decision variables. We first present a transformation technique that converts the constrained optimization problem with invalid configuration constraints into an equivalent unconstrained problem by incorporating a penalizing term into the cost function. Then, leaning upon control theory, we propose an alternative method tailored for feedback-based quantum algorithms that directly tackles IC constraints without requiring slack variables. Our approach introduces a new operator that encodes the optimal feasible solution of the constrained optimization problem as its ground state. Then, a controlled quantum system based on the Lyapunov control technique is designed to ensure convergence to the ground state of this operator. Two approaches are introduced in the design of this operator to address IC constraints: the folded spectrum approach and the deflation approach. These methods eliminate the need for slack variables, significantly reducing the quantum circuit depth and the number of qubits required. We show the effectiveness of our proposed algorithms through numerical simulations.
\end{abstract}


    
    




\end{frontmatter}




\section{Introduction} \label{S1}
Variational quantum algorithms (VQAs) are considered one of the most promising strategies tailored to the limitations of noisy-intermediate scale quantum (NISQ) devices \cite{cerezo2021variational}. VQAs have been applied across diverse domains, such as quantum chemistry, error correction, quantum machine learning, and combinatorial optimization, as detailed in the review by Cerezo et al. \cite{cerezo2021variational}. Despite their promise, VQAs face significant challenges, including selecting appropriate parameterized quantum circuits and the complexity of the non-convex classical optimization for updating circuit parameters.

Another set of hybrid quantum algorithms for the preparation of ground states of Hamiltonians that are optimization-free is gradient-based algorithms such as feedback-based quantum algorithms (FQAs) \cite{magann2022lyapunov,magann2022feedback,WFQAE,larsen2024feedback,abdul2024adaptive}, randomized adaptive quantum state preparation algorithm \cite{magann2023randomized}, anti-hermitian contracted Schr\"odinger equation \cite{smart2021quantum}, and non-variational ADAPT \cite{tang2024non}. These algorithms differ fundamentally from VQAs in that they are non-variational, and hence, they avoid the reliance on iterative, non-convex classical optimization. Instead, they rely on an update law based on the gradient estimate to update the parameters of the quantum circuit. While these algorithms similarly update the circuit parameters based on the gradient estimate, they differ in how the quantum circuit is constructed.

FQAs are inspired by control theory. They have the advantage of constructing the quantum circuit iteratively by simulating a quantum dynamical system, resulting in problem-tailored quantum circuits. They build the quantum circuit incrementally, layer-by-layer, and determine the circuit parameters based on the measurements of the qubits in the preceding layer. This approach eliminates the need for a classical optimizer and ensures a monotonic enhancement of the approximate solution as the circuit depth increases. The first FQA is the feedback-based algorithm for quantum optimization (FALQON), introduced by Magann et al., as an alternative method to VQAs to address quadratic unconstrained binary optimization (QUBO) problems \cite{magann2022lyapunov,magann2022feedback}. Since then, several improvements and advancements have been proposed for FQAs \cite{WFQAE,abdul2024adaptive,malla2024feedback,brady2024focqsfeedbackoptimallycontrolled,arai2025scalable,chandarana2024lyapunov,tang2024nonvariationaladaptalgorithmquantum,rahman2024feedbacke}. In addition, FQAs have shown to be robust against noise as demonstrated in experiments on IBM’s quantum hardware \cite{blekos2024review}.

In \cite{abdul2024feedback}, the feedback-based algorithm for quantum optimization with constraints (FALQON-C) was introduced as an extension of FALQON to solve quadratic constrained binary optimization (QCBO) problems involving equality and inequality constraints. In this work, we further expand the capabilities of FALQON-C to address a specific category of constraints known as invalid configuration (IC) constraints. These constraints explicitly prohibit specific configurations of the decision variable and can be expressed as:
$$x \notin \mathcal{E}:=\{z^{(1)},\dots,z^{(n_1)}\},$$ 
where $x$ is the decision variable and $\mathcal{E}$ is the set of invalid configurations. Such IC constraints are significant as they arise in numerous optimization problems, such as the shortest vector problem (SVP) \cite{albrecht2023variational}. In SVP, the all-zero state is prohibited ($x\neq 0$) since it corresponds to the trivial solution of the zero vector. Instead, the solution is encoded by the first excited state of the resulting Hamiltonian, which is derived by converting SVP, formulated as a QUBO problem, into an Ising Hamiltonian \cite{albrecht2023variational}.

To solve the constrained optimization problem with IC constraints using FALQON, it should first be transformed into an equivalent unconstrained problem by adding a penalizing term to the cost function to address the IC constraints. In this work, we give a general approach to convert the IC constraints into a penalizing term added to the cost function. We show that for the conversion, we need to introduce $n_1 \times (n-2)$ slack variables, where $n$ is the number of decision variables and $n_1$ is the number of IC constraints. This transformation enables the application of FALQON and FALQON-C to QCBO problems with IC constraints. However, the addition of slack variables is known to complicate the optimization procedure by increasing the search space and the complexity of the optimization landscape \cite{montanez2024unbalanced,schnaus2024efficient,hess2024effective,vyskocil2019embedding}.  Despite these challenges, this approach remains beneficial for quantum algorithms that require the problem to be in a QUBO formulation, such as the quantum approximate optimization algorithm (QAOA) and quantum annealing \cite{hauke2020perspectives}.

In this work, we propose a novel approach based on control theory for FQAs to directly address the IC constraints. Instead of converting the constrained optimization problem into an equivalent unconstrained problem, we directly tackle the constrained problem by introducing a new operator that encodes the optimal feasible solution as its ground state. We design this operator by shifting the energies of all the states associated with infeasible outcomes such that the ground state of this operator is the minimum-energy feasible solution. Subsequently, we design a Lyapunov control law to converge from all initial states to the ground state of this operator. Building on this control framework, we present the feedback-based algorithm for quantum optimization with invalid configuration constraints (FALQON-IC). When applied to QCBO problems, we show that our approach reduces the quantum resources by lowering the quantum circuit's depth and the number of qubits required, compared to FALQON and FALQON-C.

\begin{figure}[H]
      \centering
      \includegraphics[width=1\linewidth]{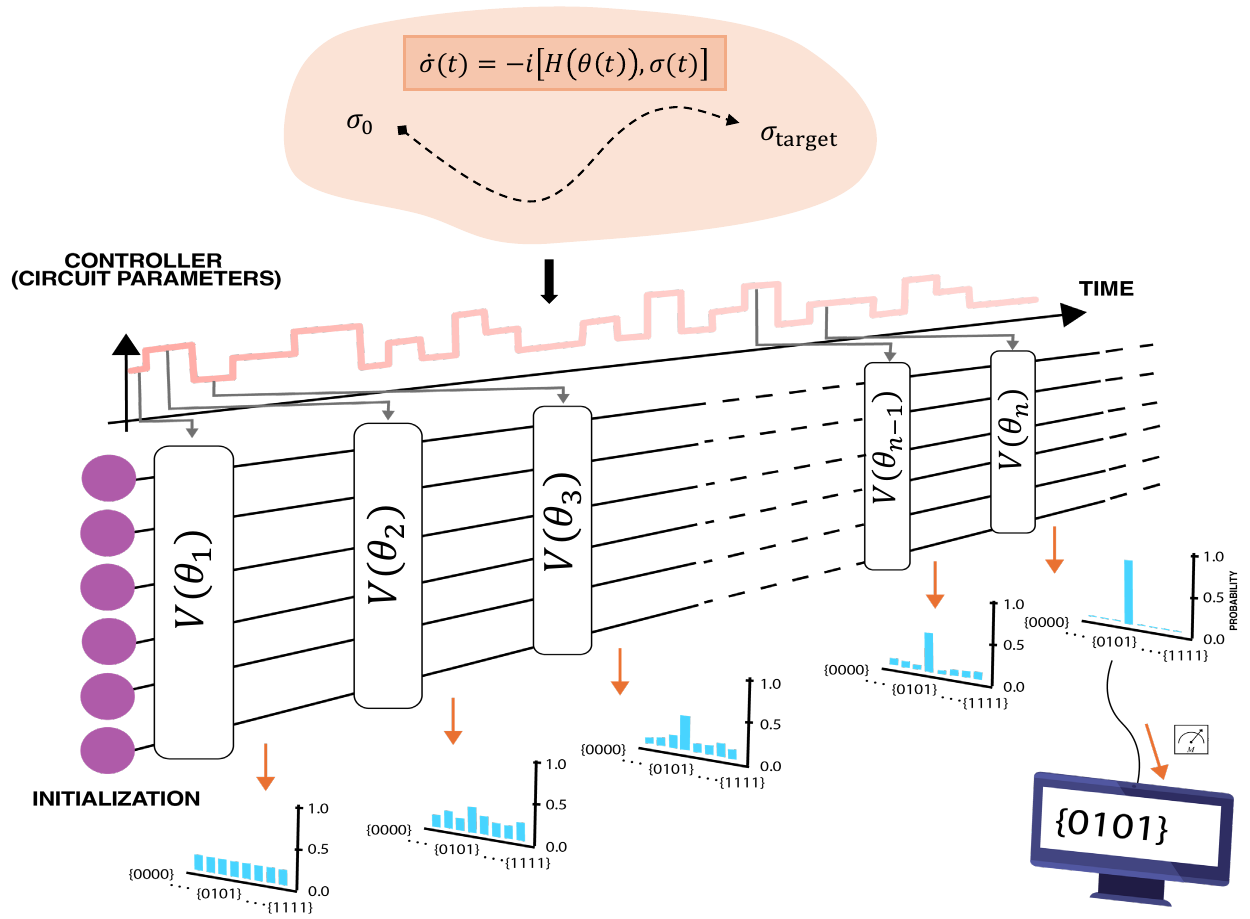}
       \caption{Illustrative diagram of feedback-based quantum algorithms. In feedback-based quantum algorithms, a controlled quantum dynamical system is designed such that the quantum states evolve from any initial state to the target state that encodes the solution to the given problem. Leveraging the Lyapunov control technique, the system minimizes the Lyapunov function (cost function) along the trajectories of the dynamical system. The problem's solution is obtained by simulating the trajectory of this dynamical system on a quantum computer, starting from an easily preparable initial state and employing a time-dependent Hamiltonian simulation algorithm. This process constructs the quantum circuit dynamically, layer-by-layer, with each layer's parameters determined through a feedback control law, computed using hybrid classical and quantum computers. This methodology enables the transformation of variational quantum algorithms into feedback-based quantum algorithms by replacing the classical optimization loop with a deterministic control law for assigning the circuit parameters. The style of the figure is inspired by \cite{hu2024overcoming} with a different content.}
      \label{G1FQA}
   \end{figure}

The design methodology for FALQON-IC leverages control theory to enhance the algorithm's efficiency. Instead of modifying the control system by introducing a new problem Hamiltonian derived from converting the equivalent QUBO problem into an Ising Hamiltonian, we alter the Lyapunov function rather than the system's problem Hamiltonian. This modification informs the system of the change in the targeted eigenstate by utilizing a newly constructed operator that encodes information about the updated target. This method enhances the quantum circuit implementation for the problem operator by eliminating additional terms introduced by the constraints. The control law is updated to incorporate the newly constructed operator, updating the system about the change in the target eigenstate. To design the operator within the Lyapunov function, we explore two techniques: deflation methods \cite{mackey2008deflation} and the folded spectrum (FS) approach \cite{cadi2024folded}. Deflation techniques have been utilized to develop efficient quantum algorithms for preparing excited states of Hamiltonians \cite{rahman2024feedbacke, higgott2019variational}. The folded spectrum method has also demonstrated its utility in designing quantum algorithms for quantum chemistry and combinatorial optimization problems \cite{cadi2024folded, bauer2024combinatorial}. In this work, we adapt these methods to efficiently handle IC constraints, enabling practical and scalable solutions for QCBO problems. A schematic representation of our approach is shown in Figure \ref{G1FQA}.

We highlight the main contributions of this work as follows: 
\begin{itemize}

    \item We propose a general discrete model for the quantum system, replacing the continuous model, and apply discrete-time Lyapunov control directly at the circuit level. This discrete model offers the advantage of being directly realizable as quantum gates, offering a layer of abstraction separating from the continuous model, which is related to the pulse level. For further insights and discussions regarding optimization on the circuit level, refer to \cite{magann2021pulses}.
    
    \item We propose a general framework for encoding IC constraints as a penalty term within the cost function. This approach is particularly beneficial for algorithms that require the problem to be in a QUBO formulation, including QAOA and quantum annealing \cite{hauke2020perspectives}. 
       
    \item We introduce a novel feedback-based quantum algorithm to solve QCBO problems with equality, inequality and IC constraints, namely FALQON-IC. The algorithm utilizes folded spectrum and deflation techniques to tackle the IC constraints directly. We give general approaches to tuning the hyperparameters to ensure the encoding of the problem's solution. Our approach substantially reduces the quantum resources required to implement the quantum circuit for the algorithm.

    \item We give a thorough overview of existing FQAs for solving QCBO problems, including FALQON, FALQON-C, and our newly proposed algorithm, FALQON-IC. These algorithms rely on a feedback-based parameter update law to assign the circuit parameters. 
    
\end{itemize}
The rest of the paper is organized as follows. Section 2 provides an overview of QCBO problems and a general description of the Lyapunov control of quantum systems in both continuous-time and discrete-time frameworks. Section 3 outlines a general method for converting QCBO problems into equivalent QUBO problems. Section 4 introduces FALQON-IC, our primary algorithm for addressing IC constraints, utilizing deflation and folded spectrum approaches. Section 5 presents an estimation of the quantum resources required for implementing the proposed algorithms and demonstrates the effectiveness and scalability of our approach through numerical simulations. Section 6 concludes the paper and discusses directions for future work.

\section{Preliminaries}  \label{S2}

This section begins by formulating the general QCBO problem, including the IC constraints. Next, we provide an overview of Lyapunov control theory for continuous-time and discrete-time quantum systems, which is the foundation for our general feedback-based quantum algorithm framework.

\subsection{Quadratic Constrained Binary Optimization Problem} \label{ssec2}
The general QUBO problem is given as follows:
\begin{equation} \label{QUBO}
    \min _{x \in \{0,1\}^n} J(x) := x^TT_Jx+c_J^Tx+a_J
\end{equation}
Here, $T_J \in \mathbb{R}^{n\times n}$ is a symmetric matrix, $c_J$ $\in \mathbb{R}^{n}$ and $a_J$ $\in \mathbb{R}$. FALQON was originally proposed to tackle QUBO problems \cite{magann2022lyapunov,magann2022feedback}. In \cite{abdul2024feedback}, FALQON-C was introduced to efficiently extend FALQON for tackling QCBO with equality and inequality constraints in the following general form:
\begin{equation} \label{constraint}
    G^{(q)}(x) := x^TT_{G}^{(q)}x+c_{G}^{(q)T} x+a^{(q)}_{G} \star 0, \;\;\; q=1, 2,3, ...,n_2,
\end{equation}
where we use $\star$ to denote a comparison operator $\star \in \{\leq, = \}$. The matrices $T_{G}^{(q)} \in \mathbb{R}^{n\times n}$ are symmetric, $c_{G}^{(q)}$ $\in \mathbb{R}^{n}$ and $a_{G}^{(q)}$ $\in \mathbb{R}$. In this work, we show how feedback-based quantum algorithms can be efficiently extended to tackle QCBO with IC constraints given in the following form:
\begin{equation} \label{ICs}
    x \notin \{z^{(1)}, z^{(2)}, \dots, z^{(n_1)} \}\subset \{0,1\}^{n}, 
\end{equation}
where $z^{(r)} \in \{0,1\}^n$ is a specific invalid configuration. Hence, the general form of the QCBO becomes
\begin{subequations}
	\label{QCBO}
	\begin{align}
		&\min _{x \in \{0,1\}^n} J(x) = x^TT_Jx+c_J^Tx+a_J\\
		&\text{s.t.}  
         \; \quad\;  x \neq z^{(r)}, \quad \;\;\; r=1, 2,3, ...,n_1 \\ &
        \quad G^{(q)}(x) = x^TT_{G}^{(q)}x+c_{G}^{(q)T} x+a^{(q)}_{G} \star 0, \;\;\; q=1, 2,3, ...,n_2. \label{IC}
	\end{align}
\end{subequations}

\subsection{Continuous-Time Quantum Lyapunov Control} \label{SS2.1}
Let $ \mathcal{H} = \mathbb{C}^N $ represent the state space with an orthonormal basis $ \mathcal{B} = \{ |0\rangle,\ket{1}, \dots, \ket{N-1} \} $. The space of density operators is denoted by $ \mathcal{D} = \{ \sigma \in \mathbb{C}^{N \times N} \mid \sigma \geq 0, \, \text{Tr}(\sigma) = 1 \} $. Henceforth, we will represent all operators on the $ \mathcal{B} $ basis.

Consider a quantum system described by the controlled master equation:
\begin{equation} \label{master}
    \dot{\sigma}(t)=-i\left[H_{P}+\theta(t) H_{M}, \sigma(t)\right], \quad \sigma(0)=\sigma_{0},
\end{equation}
where we normalize the Plank constant to $\hbar=1$, the control input is $\theta (t)$, the \textit{problem Hamiltonian} $H_P$ has eigenvectors $\{\ket{e_0},\ket{e_1}, \dots, \ket*{e_{N-1}}\}$ and corresponding eigenvalues $\{e_0>e_1> \dots > e_{N-1} \}$ and $H_M$ is the \textit{mixer Hamiltonian}. This work assumes that the Hamiltonians $H_P$ and $H_M$ are time-independent and non-commuting, i.e. $[H_P,H_M] \neq 0$. While we focus on a single control input for simplicity, the analysis can be extended to multiple control inputs (see \ref{appA.1} for details).

We define the operator $Q$ that commutes with $H_P$ ($[Q,H_P]=0$) and have eigenvalues $\{\zeta_0>\zeta_1> \dots > \zeta_{N-1} \}$. Our goal is to find the ground state of this operator, defined as $\sigma_g := \arg \min_{\sigma \in \mathcal{D}} \operatorname{Tr}(Q \sigma)$. To achieve this, we aim to design a feedback law in the form $\theta = \Lambda(\operatorname{Tr}(A \sigma))$, where $A$ is an observable and $\Lambda(\cdot): \mathbb{R} \to \mathbb{R}$ is a general function of the expectation value of $A$, which stabilizes the state $\sigma$ at $\sigma_g$. Choosing this feedback form facilitates its evaluation using the quantum computer.

Let the Lyapunov function be
\begin{equation}
    L(\sigma) = \operatorname{Tr}(Q \sigma).
\end{equation}
The derivative of this Lyapunov function along trajectories is given as follows:
\begin{equation}
    \nabla L \cdot \dot{\sigma}= \operatorname{Tr}\left(i\left[H_M,Q\right] \sigma(t)\right) \theta(t) .
\end{equation}
To guarantee that the Lyapunov function is non-increasing, i.e. $ \nabla L \cdot \dot{\sigma} \leq 0$, we design the controller $\theta(t)$ in the following way:
\begin{equation} \label{controller1c}
    \theta(t)=- \kappa \Lambda \bigg(\operatorname{Tr} \big(i\left[H_M,Q\right] \sigma(t)\big) \bigg),
\end{equation}
where $\kappa>0$ is the controller gain and $\Lambda(\cdot): \mathbb{R} \to \mathbb{R}$ is a function that satisfies the condition $\alpha \Lambda(\alpha) > 0$ $\forall \alpha \neq 0$. The graph of the function $\beta = \Lambda(\alpha)$ passes through the origin on the $\alpha - \beta$ plane monotonically and lies entirely within Quadrant I or Quadrant III, ensuring that the given condition is satisfied. For the rest of the analysis, we will choose $\Lambda(\cdot)$ as the identity function, i.e. $\Lambda(\alpha)=\alpha$. Alternative design choices for $\Lambda(\cdot)$ are discussed in \ref{appB}. For this design of the controller, we get:
\begin{equation} \label{Vdot}
  \nabla L \cdot \dot{\sigma} = -\kappa \big|\operatorname{Tr}\left(i\left[H_M,Q\right] \sigma(t)\right)\big|^2 \leq 0 .
\end{equation}
The application of the controller \eqref{controller1c}, and based on specific assumptions (outlined in Section 2.1 of \cite{clausen2023measurement}), guarantees asymptotic convergence to the ground state $\sigma_g$ of the operator $Q$. We summarize the design principles of the operator $Q$ as follows \cite{clausen2023measurement}:
\begin{itemize}
    \item  The designed operator $Q$ should commute with the problem Hamiltonian $H_P$, i.e. $[Q,H_P]=0$.
    \item The operator $Q$ should have a non-degenerate spectrum, i.e. $\zeta_q \neq \zeta_r$ for all $q \neq r$.
    \item The eigenvalue $\zeta_g$ associated with the target eigenstate must be the smallest among all eigenvalues, indicated by $\zeta_g < \zeta_q$ for $q \in (0,1,...,M-2)$ where $\zeta_q \neq \zeta_g$.
\end{itemize}

The time evolution operator associated with the master equation \eqref{master} is given as $V(t_0, t) = \chi e^{-i \int_{t_0}^{t} H(x) dx}$ with $\chi$ being the time-ordering operator. Assuming the function $\theta (t)$ to be piecewise-constant in small intervals $\Delta t$, we decompose this operator into $r$ piecewise-constant intervals of a small length of $\Delta t$ to get:
\begin{align}
V(0, T) = \chi e^{-i \int_{0}^{T} H(x) dx} \approx \prod_{n=1}^{r} e^{-i\left(\theta(n \Delta t) H_{M} + H_{P}\right) \Delta t},
\end{align}
where  $T=r\Delta t$. Moreover, since the Hamiltonians $H_P$ and $H_M$ do not commute ($[H_M,H_P]\neq0$), we employ Suzuki-Trotter first order approximation. Hence, we get:
\begin{align}
V(0, T)  \approx \prod_{n=1}^{r} e^{-i \theta(n \Delta t)  H_{M}\Delta t}  e^{-iH_{P}\Delta t} = \prod_{n=1}^{r} V_M(\theta_n)   V_P,
\label{evolution}
\end{align}
where $\theta_n=\theta(n \Delta t)$, $V_P=e^{-iH_P\Delta t}$, and $V_M(\theta_n)=e^{-i \theta(n \Delta t)  H_M\Delta t}$. By choosing $\Delta t$ to be sufficiently small, it can be guaranteed that condition \eqref{Vdot} is satisfied (see \ref{appA.2} for details). For the control law, we use a discrete version of the controller from \eqref{controller1c}, as shown below:
\begin{equation}
    \theta_{n+1}= - \kappa \operatorname{Tr} (\left[H_M,Q\right] \sigma_n),
    \label{udis}
\end{equation}
where  $\sigma_n=\sigma(n \Delta t)$,
Note that the controller at each discrete time step corresponds to the circuit parameters, and in this work, the terms \textit{controller} and \textit{circuit parameters} will be used interchangeably. 
\subsection{Discrete-Time Quantum Lyapunov Control} \label{ss2.2}
Rather than the continuous-time system \eqref{master}, we consider the following discrete-time formulation:
\begin{equation} \label{model2}
    \sigma_{n+1}=V\left(\theta_n\right) \sigma_n V\left(\theta_n\right)^{\dagger},
\end{equation}
where $n =  1,2,3,\dots$ denotes the discrete-time index. The operator $V(\theta)$ is a unitary operator parameterized by the control input and is defined as $V(\theta)=V_PV_M(\theta)$ with $V_M(\theta)=e^{-i\theta H_M\Delta t }$ and $V_P=e^{-iH_P \Delta t}$. As shown in Subsection~\ref{SS2.1}, this formulation approximates the continuous-time master equation using a piecewise-constant discretization. The discrete-time system offers the advantage of being directly implementable as a quantum circuit, offering a layer of abstraction separating from the continuous model (see \cite{magann2021pulses} for additional discussions).

As for the continuous case, we define the operator $Q$ that commutes with $H_P$ ($[Q,H_P]=0$) and design a feedback law in the from $\theta=\Lambda(\operatorname{Tr}(A \sigma))$ that stabilizes the state $\sigma$ at $\sigma_g$. Let the Lyapunov function be
\begin{equation}
    L(\sigma) = \operatorname{Tr}(Q \sigma).
\end{equation}
The aim is to guarantee that the Lyapunov function is non-increasing, thus satisfying $L(\sigma_{n+1})-L(\sigma_n) \leq 0$.
Using first-order Taylor series expansion, we expand the operator $V_M(\theta)$ as follows:
\begin{equation} \label{T1}
    V_M(\theta)=e^{-i\theta H_M\Delta t} = I - i\Delta t H_m \theta + O(\Delta t^2),
\end{equation}
We get:
\begin{align} \label{DV}
 L(\sigma_{n+1})-L(\sigma_n)  &= \operatorname{Tr} \big(QV_PV_M(\theta_n) \sigma_n V_M(\theta_n)^\dagger V_P^\dagger \big) - \operatorname{Tr}(Q\sigma_n) \nonumber \\
&= \theta_n \Delta t\operatorname{Tr}(i[H_M,Q]\sigma_n)+O(\Delta t^2).
\end{align}
To guarantee that the Lyapunov function is non-increasing, i.e. $L(\sigma_{n+1})-L(\sigma_n) \leq 0$, we choose $\Delta t$ to be sufficiently small (see \ref{appA.2}) and design the controller in the following way:
\begin{equation} \label{controller1}
    \theta_{n+1}=-\kappa \Lambda \bigg(\Delta t\operatorname{Tr} \big(i\left[H_M,Q\right] \sigma_n\big) \bigg),
\end{equation}
where $\kappa>0$ is the controller gain. The application of the controller \eqref{controller1}, and based on specific assumptions (outlined in Section 2.1 of \cite{clausen2023measurement}), guarantees asymptotic convergence to the ground state $\sigma_g$ of the operator $Q$.

The Lyapunov control framework presented in this section extends beyond the original formulation used in FALQON \cite{magann2022lyapunov,magann2022feedback}. In FALQON, the operator $Q$ is simply chosen as the problem Hamiltonian $H_P$, relying on the fact that the ground state of $H_P$ encodes the solution. In our previous work \cite{abdul2024feedback}, we utilized this general Lyapunov control framework formulation to develop FALQON-C, where $Q$ was specifically designed to encode the solution of QCBO problems with equality and inequality constraints in its ground state. In this work, we leverage this framework to introduce FALQON-IC, which further extends FALQON-C to handle IC constraints.

In Section \ref{S3}, we show how we can convert IC constraints into equivalent penalizing term in the cost function using slack variables. Subsequently, in Section~\ref{S4}, we propose an innovative method to extend FALQON-C for addressing problems with IC constraints without requiring slack variables. This is achieved by designing an appropriate Lyapunov function and employing deflation and folded spectrum techniques.

\section{Invalid-Configuration Constraints}\label{S3}
In this section, we introduce a general approach for addressing IC constraints in the general form of $x \neq z$ where $z$ can be any invalid configuration. This approach converts the IC constraint into an equivalent penalization term added to the cost function. A special case of this problem was previously addressed in the context of the SVP with a constraint of the form $ x \neq 0 $ \cite{albrecht2023variational}. In this work, we extend this method to accommodate arbitrary IC constraints.

To solve the QCBO problem given by \eqref{QCBO} using FALQON, it should first be converted into an equivalent QUBO problem. In this work, the equivalence of the problems is defined as follows. \\

\noindent
\textbf{Definition 1:} Two problems, \(A\) and \(B\), are considered equivalent if their sets of optimal solutions, denoted as \(\mathcal{X}^*_A \subseteq \mathcal{X}\) and \(\mathcal{X}^*_B \subseteq \mathcal{X}\), are identical, i.e., \(\mathcal{X}^*_A = \mathcal{X}^*_B\). \\

We present a general approach for achieving this conversion by introducing a penalty term into the cost function to account for the IC constraints. This conversion requires the addition of $n_1 \times (n-2)$ slack variables, where $n$ is the number of decision variables and $n_1$ is the number of IC constraints. In this approach, we aim to penalize outcomes that correspond to IC constraints. This is achieved by modifying the objective function to include a penalty term. The penalty is constructed such that it evaluates to zero for feasible outcomes while assigning sufficiently large positive values to outcomes associated with IC constraints.

Without loss of generality, let us assume that there is only one invalid configuration $z$.
Suppose we want to convert the IC constraint into a penalizing term in the cost function. In that case, we can use the following penalizing term added to the cost function to penalize the invalid configurations $z$.
\begin{equation} \label{HOconst}
    \gamma \prod_{q=1}^n 
    \overline{x_q\oplus z_q},
\end{equation}
where $\gamma$ is a hyperparameter chosen large enough to ensure that the resulting problem is equivalent to the QCBO problem, $\oplus$ denotes exclusive OR modulo 2 addition, and the overline is logical negation. Note that the penalty term in \eqref{HOconst}  can also be written as 
\begin{equation} \label{HOconst2}
    \gamma \prod_{q=1}^n \Big(1 -\big(z_q(1-x_q) + (1-z_q) x_q \big)\Big),
\end{equation}
which results in higher-order terms, rendering the problem no longer quadratic. Hence, we will propose an equivalent penalizing term that results in a QUBO formulation of the problem. 

Let us define $h: \mathbb{Z}_2^n \to \mathbb{Z}_2^n$ by
\begin{equation} \label{h(x)}
    h(x):=x\oplus z,
\end{equation}
where $\oplus$ acts element-wise. For the case of $n$ decision variables, we add $n-2$ binary slack variables $s=\{s_q\}_{q = 1,2, \dots, n-2}$ and let $v(x,s)=[s_1,s_2, \dots, s_{n-2},1-h_n(x),1]$,  where $h_n$ refers to the $n$th output of $h$, namely $h_n(x)=x_n\oplus z_n$. To suppress the notation, we write $h$ for $ h(x) $ and $v$ for $v(x,s)$. Next, consider the upper-triangular matrix
\begin{equation} \label{A}
 A = \begin{bmatrix}
-1 & 1 & 1 & \cdots & 1 \\
0 & -1 & 1 & \cdots & 1 \\
0 & 0 & -1 & \cdots & 1 \\
\vdots & \vdots & \vdots & \ddots & \vdots \\
0 & 0 & 0 & \cdots & -1
\end{bmatrix}.
\end{equation}
Finally, define the function $g: \mathbb{Z}_2^{n} \times \mathbb{Z}_2^{n-2} \to \mathbb{Z} $ as follows:
\begin{equation} \label{g}
    g(x,s)=\tilde g(h(x),v(x,s)) = 1 + v^T A h. 
\end{equation}
Later, in Proposition~1, we will show that $g$ is non-negative; it evaluates to one for IC constraints, and for an appropriate choice of the slack variables $s$, it can be made zero for the feasible states. Therefore, the function $g$ can be used as a penalty term in the cost function, which will be shown in Theorem~1.

Now, we provide a theorem to set an equivalence between optimization problems with and without IC constraints, excluding the case of inequality constraints since this is well-known in literature such as \cite{glover2022quantum}. \\ \\
\noindent
\textbf{Theorem 1.} Let $\mathcal{E}$ be the set of infeasible outcomes due to the IC constraints defined as $\mathcal{E}=\{z^{(1)},\dots,z^{(n_1)}\}$. Let problem A be formulated as follows:
\begin{subequations}
	\label{QCBOT}
	\begin{align}
		&\min _{x \in \{0,1\}^n} F(x) := x^TT_Fx+c_F^Tx+a_F\\
		&\text{s.t.} \nonumber \\
           & \; \quad \quad  x \notin \mathcal{E}, \label{ICT}
	\end{align}
\end{subequations}
where $T_F\in \mathbb{R}^{n\times n}$ is a symmetric matrix, $c_F$ $\in \mathbb{R}^{n}$ and $a_F$ $\in \mathbb{R}$. 
We denote solution of Problem~A by $\mathbf{x}_A^*=\arg \min _{{x} \in\{0,1\}^n \backslash  \mathcal{E}} F({x})$.\\
Let problem B be formulated as follows:
\begin{equation}
    \min _{y \in \{0,1\}^{n+n_1(n-2)}} \bar F(y) :=  y^ T \bar T_Fy+ \bar c_F^Ty+  a_F + \sum_{r=1}^{n_1} \gamma_r g^{(r)}(y),
\end{equation}
where  $y=(x, s)$ and $s=(s_1^{(1)}, \dots, s_{n-2}^{(1)}, \dots, s_1^{(n_1)}, \dots, s_{n-2}^{(n_1)})$ is the vector of added slack variables, $$
  \bar T_F = 	\begin{bmatrix}
T_F & 0_{n \times n_1(n-2)} \\
0_{n_1(n-2) \times n} & 0_{n_1(n-2) \times n_1(n-2)}
\end{bmatrix}, 
$$ 
$$
\bar c_F = \begin{bmatrix}
c_F^T  & 0_{1\times n_1(n-2)} 
\end{bmatrix}^T,
$$ 
$\{\gamma_r\}_{r = 1,2, \dots, n_1}$ are shifting parameters chosen such that \\$\gamma_r > \max{\{0,  F(\mathbf{x}_A^*) - F(z^{(r)})\}}$,
and $g^{(r)}$ is given as follows:
\begin{equation}
    g^{(r)}(y):=g^{(r)}(x,s):=\tilde g^{(r)}(h^{(r)}(x),v^{(r)}(s,x)) = 1 + {v^{(r)}}^T A h^{(r)},
\end{equation}
where $h^{(r)}(x):=x\oplus z^{(r)}$, $v^{(r)}(x,s):=[s^{(r)}_1,s^{(r)}_2, \dots, s^{(r)}_{n-2},1-h^{(r)}_n,1]$, and $A$ is as given in \eqref{A}. Here, $O_{q \times j}$ represents a zero matrix of dimensions $q \times j$. A solution of problem B is $\mathbf{x}_B^*$, where $(\mathbf{x}_B^*,s^*)=\mathbf{y}_B^*=\arg\min _{\mathbf{y} \in\{0,1\}^{n+n_1(n-2)}} \bar F(\mathbf{y})$, i.e. $\mathbf{x}_B^*$ is the value of the projection of $\mathbf{y}_B^*$ on the first n variables. Then, problems A and B are equivalent. 
\\
\\
Proof: Without loss of generality, we assume that the set $\mathcal{E}$ is a singleton, $\mathcal{E}=\{ z \}$. The proof of Theorem 1 leans upon the following result. 
\begin{proposition}
The following statements hold:
\begin{itemize}
\item[(P1)] $\forall x,s \ g(x,s)\geq 0$ (or equivalently
$\bar F(x,s)\geq F(x)$),
\item[(P2)] $\forall s \ g(z,s)=1$ (or equivalently $\bar F(z,s)=F(z)+\gamma$),
\item[(P3)] $\forall x\neq z \ \exists s^*\ g(x,s^*)=0$ (or equivalently $\bar F(x,s^*)=F(x)$).
\end{itemize}
\end{proposition}

Proof of P1:
Let $q_\text{max}$ be the largest $q$ for which $h_q=1$ (equivalently $x_q\neq z_q$). It can be seen that $$Ah=[\ *\ \cdots\ *\underbrace{-1}_{q_\text{max} th\ \text{term}}0\ \cdots\ 0\ ]^T$$ where $*$'s are some non-negative numbers. Since $v$ is also a binary vector, $v^TAh\geq -1$ and hence $g(x,s)\geq 0$. 

Proof of P2:
Recall from \eqref{h(x)} that $h(x)= x \oplus z$ with $\oplus$ operating element-wise, and hence when $x=z$, $h$ is the zero vector and consequently $g(z,s)=1$ for all $s$. 

Proof of P3: Assume that $x \neq z $. If $q_\text{max}=n$ then $h_n=1$, $Ah=[ *,\cdots,*,-1]$ and setting $s_q^*=0$ for all $q=1,\dots,n-2$ leads to $v=[0, \dots,0,1]$ and hence to $ g(x,s^*)=\tilde g(h(x),v(x,s^*)) =g(v,h)=1 + v^T A h=0$. If $q_\text{max}=n-1$ then $h_{n-1}=1$, $h_n=0$, $Ah=[ *,\cdots,*,-1,0]$ and setting $s_q^*=0$ for all $q=1,\dots,n-2$ leads to $v=[0, \dots, 0,1,1]$ and hence to $g(x,s^*)=0$. Finally, if $q_\text{max}<n-1$, then setting $s_{q_\text{max}}^*=1$ and $s_q^*=0$ for all $q\neq q_\text{max}$ leads to 
$$v=[\ 0\cdots 0 \underbrace{1}_{q_\text{max} th\ \text{term}} 0\ \cdots\ 0\ 1]^T$$ and hence $g(x,s^*)=0$. $\Box$

Proof of Theorem~1: 
Note that $\bar F(y) = \bar F(x,s) = F(x) + \gamma g(x,s)$.

Suppose that $\mathbf{x_A^*}$ is a solution of problem A. Then $\textbf{x}_A^*\neq z$ and $x\neq z$ implies $F(x)\geq F(\textbf{x}_A^*)$. We will show that exist slack variables $s^*$ such that $(\mathbf{x_A^*}, s^*)$ is a solution of problem B, or equivalently $\bar F(x,s)\geq \bar F(\textbf{x}_A^*,s^*)$ for all $x$ and $s$. 
Since $\textbf{x}_A^*\neq z$, by P3, there exists $s^*$ such that $\bar F(\textbf{x}_A^*,s^*)=F(\textbf{x}_A^*)$.
If $x\neq z$, then by P1 and by the fact $x\neq z$ implies $F(x)\geq F(\mathbf{x_A^*})$, we get $\bar F(x,s)\geq F(\mathbf{x_A^*}) = \bar F(\mathbf{x_A^*},s^*)$. 
If $x=z$, then by P2, we get  $\bar F(x,s)=\bar F(z,s)=F(z)+\gamma$ and hence, by the assumption that $\gamma > \max{\{0,  F(\mathbf{x_A}^*) - F(z)\}}$, we get $\bar F(x,s) > F(\mathbf{x_A^*}) = \bar F(\mathbf{x_A^*},s^*)$.

Suppose that $\mathbf{x_B^*}$ is a solution of problem B. Then there exists $s^*$ such that  $y_B^*=(\textbf{x}_B^*,s^*)=\arg\min _{\mathbf{y} \in\{0,1\}^{2n-2}} \bar F(\mathbf{y})$, equivalently, $\bar F(x,s)\geq \bar F(\textbf{x}_B^*,s^*)$ for all $x$ and $s$. We will show that $\mathbf{x_B^*}$ is a solution of problem A. In other words, we will show that $\textbf{x}_B^*\neq z$ and $x\neq z$ implies $F(x)\geq F(\textbf{x}_B^*)$. Note that P2 and the assumption that $\gamma_r > \max{\{0,  F(\mathbf{x_A}^*) - F(z^{(r)})\}}$ implies $\bar F(z,s)>F(\mathbf{x}_A^*)$. Hence $\textbf{x}_B^*\neq z$ follows. If also $x\neq z$ then by P3, there exists $\hat s$ such that $F(x)=\bar F(x,\hat s)$ and since $\textbf{x}_B^*$ is a solution of problem B, we get $F(x)\geq \bar F(\mathbf{x_B^*},s^*)$ for some $s^*$. Now P1 for $x=\mathbf{x_B^*}$ and $s=s^*$ implies that $F(x)\geq F(\mathbf{x_B^*})$. $\Box$

From Theorem 1, the QCBO with IC constraints can be converted into an equivalent QUBO problem, and hence, FALQON \cite{magann2022lyapunov} can be adapted to solve the equivalent QUBO problem. However, as previously demonstrated, this approach requires introducing $n_1\times(n-2)$ slack variables, drastically increasing the number of qubits. To address this limitation, the next section presents FALQON-IC for solving QCBO problems by designing a tailored observable for the Lyapunov function. \\

\begin{remark}
     According to Theorem 1, it is necessary to select $\gamma_r > \max{\{0,  F(\mathbf{x_A}^*) - F(z^{(r)})\}}$ for $r \in \{1,2,\dots,n_1\}$. While we can evaluate $F(z^{(r)})$, $\mathbf{x_A}^*$ is not known in advance, and hence, we cannot evaluate $F(\mathbf{x_A}^*)$.  Later in Subsection~\ref{ss.Df}, we will show different approaches to selecting a reasonable value for $\gamma_r$ without prior knowledge of $\mathbf{x_A}^*$.
\end{remark}

\begin{remark}
An alternative approach for addressing IC constraints is to reduce the higher-order terms in \eqref{HOconst} into quadratic terms. However, this method typically involves introducing additional slack variables, which can substantially increase the problem's complexity and render the problem computationally intractable \cite{dattani2019quadratization}. 
\end{remark}

\section{Feedback-Based Quantum Optimization with Invalid Configuration Constraints} \label{S4}
This section proposes various strategies for solving QCBO problems with equality, inequality, and IC constraints. The case of equality and inequality constraints was addressed in our previous work \cite{abdul2024feedback}. We now focus on handling IC constraints. We propose three approaches: FALQON-C utilizing the conversion introduced in Theorem 1, FALQON-IC with the deflation approach, and FALQON-IC with the folded spectrum approach. These approaches are used to design a Lyapunov function that incorporates an operator specifically designed to encode the solution to the QCBO problem as its ground state. First, in Subsection~\ref{ssFC}, we review FALQON-C, which addresses QCBO problems with equality and inequality constraints but does not consider IC constraints. Then, in Subsection 4.2, we show how the conversion introduced in Theorem~1 can be utilized in FALOQN-C to address IC constraints. Following that, in Subsections 4.3 and 4.4, we introduce FALQON-IC with the deflation and folded spectrum approaches, respectively.

\subsection{Feedback-Based Algorithm for Quantum Optimization with Constraints} \label{ssFC}
In this subsection, we review FALQON-C for addressing QCBO with equality and inequality constraints as introduced in \cite{abdul2024feedback}. Subsequently, in Subsection 4.2, we show how FALQON-C can address IC constraints using the conversion introduced in Theorem~1.

In FALQON-C, the operator $Q_c$ is designed to encode the solution of the QCBO in its ground state. The procedure to design this operator is given in the following. First, the cost function $J$ is converted into a problem Hamiltonian that satisfies $H_P\ket{x}=J(x)\ket{x}$. This mapping is achieved by mapping each binary variable $x_q$ in the cost function of \eqref{QUBO} to a Pauli-Z operator using the transformation $x_q \mapsto \frac{1}{2}(I-Z_q)$, where $Z_q$ is the Pauli Z operator applied to the $q$th qubit. This results in the following problem Hamiltonian \cite{hadfield2021representation}: 
\begin{align}
\label{Hc}
H_P = \sum_{q,j=1}^{n}\frac{1}{4} T_{J,q,j}Z_qZ_j - \sum_{q=1}^{n} \frac{1}{2}\big(c_{J,q}+ \sum_{j=1}^{n}T_{J,q,j}\big)Z_q,
\end{align}
where the identity offset term $\left(\sum_{q,j=1}^{n}\frac{1}{4} T_{J,q,j}+\sum_{q=1}^{n}\frac{1}{2} c_{J,q}+a_J\right)I$ is discarded since it does not affect the solution of the problem.

Next, the inequality constraints $G^{(q)}$ are converted into equality constraints. These, along with the other equality constraints, are then transformed into \textit{constraint Hamiltonians} $H_C^{(q)}$, where $q \in \{1, 2, \dots, n_2\}$. This conversion involves 
defining a penalty function $P^{(q)}(x)$ as $P^{(q)}(x) = |G^{(q)}(x)|^m$ and
mapping each binary variable $x_q$ to a Pauli-Z operator using the transformation $x_q \mapsto \frac{1}{2}(I - Z_q)$. Consequently, the resulting Hamiltonians satisfy $H_C^{(q)}\ket{x} = P^{(q)}(x)\ket{x}$. To ensure that the eigenvalues corresponding to the infeasible outcomes are positively shifted to higher values, it is necessary to satisfy $P^{(q)}(x) \geq 0$ for all $x$. In this work, we select $m = 2$, which leads to $P^{(q)}(x) = (G^{(q)}(x))^2$. The resulting operator is given as follows:
\begin{equation}
    Q_c := H_P+\sum_{q=1}^{n_2} \beta_q H_C^{(q)} ,
    \label{Qc}
\end{equation}
where $\{\beta_q\}_{q=1, \dots, n_2}$ are hyperparameters chosen large enough to ensure that the ground state of the operator $Q_c$ encodes the optimal feasible solution of the QCBO problem. From \eqref{Qc}, it follows that the operator $Q_c$ shares the same eigenvectors as $H_P$, satisfying the commutation relation $[H_P, Q_c] = 0$. Define the set of feasible outcomes as $\mathcal{F} = \{x \in \{0,1\}^n : P^{(q)}(x)=0, \; q\in \{1,2, \dots, n_2 \} \}$ and let us call $\mathcal{F}^c=\{0,1\}^n\setminus \mathcal{F}$ as the set of infeasible outcomes. Consequently, the eigenvalues of the operator $Q_C$ are expressed as:
\begin{equation}
\zeta_j(Q_c)= 
\begin{cases}
   e_j  & \text{for } j \in \mathcal{F} \\
   e_j+ \sum_q \beta_q P^{(q)}(j),  & \text{for } j \in \mathcal{F}^c
\end{cases}
\end{equation}
Therefore, to guarantee that the smallest eigenvalue of $Q_c$ corresponds to an eigenvector that encodes the solution to the problem \eqref{QCBO}, we need to choose $\beta_q$ to satisfy the following condition: 
\begin{align}
      \sum_q \beta_q P^{(q)}(j) & \geq e^\text f_{\min} - e_{\min}, \quad j\in \mathcal{F}^c,
\end{align}
where $e_{\min}$ and $e^\text f_{\min}$ are the minimum eigenvalue of $H_P$ and the smallest eigenvalue of $H_P$ corresponding to an eigenvector encoding a feasible outcome, respectively. We define the energy gap $e_g := e_{\max} - e_{\min} > e^\text f_{\min} - e_{\min}$ with $e_{\max}$ representing the largest eigenvalue of $H_P$. Given the problem Hamiltonian expanded in the Pauli basis as $H_P=\sum_{r=1}^{m_0} c_r O_r$, where $O_r$ are Pauli strings, , the upper bound on $e_g$ is given by: $e_g \leq 2||H_P|| \leq 2 \sum_r \abs{c_r} $. 
Without loss of generality, we assume that for each $j \in \mathcal{F}^c$, there exists at least one $q$ for which $P^{(q)}(j) \geq 1$. This can be ensured by multiplying the equality constraint by the least common multiple of the denominators of the coefficients. Setting each shifting parameter $\beta_q$ larger than $e_g$, we obtain:
\begin{align}
     \sum_q \beta_q P^{(q)}(j) & \geq e_g \sum_q P^{(q)}(j)  \geq e_g > e^\text f_{\min} - e_{\min}, \quad j\in \mathcal{F}^c,
\end{align}
which ensures that the eigenvalues corresponding to infeasible outcomes are shifted above those encoding feasible outcomes. Note that we can also find an approximate estimate of $e_g$ by estimating $e_{\min}$ and $e_{\max}$ using FALQON, for example, by estimating the ground state of $H_p$ for the former and the ground state of $-H_p$ for the latter. While this approach requires additional computations, it results in a smaller shift than the upper bound, which is preferable for improving the algorithm's performance, as suggested by numerical simulations.

\begin{figure}[H]
      \centering
       \captionsetup{justification=centering}
      \includegraphics[width=1 \linewidth]{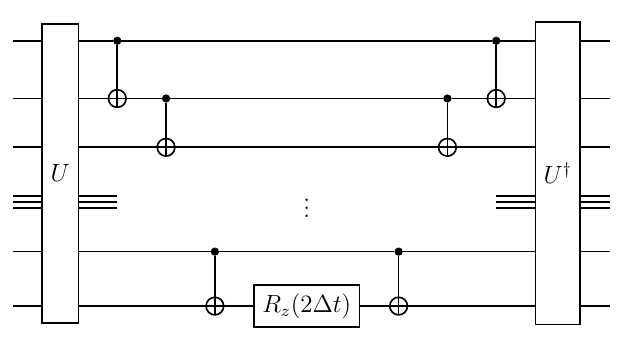}
      \caption{Quantum circuit implementation for the unitary in the general form $e^{-iO_{q} \Delta t}$, where $O_{q}=O_{q,1} \otimes O_{q,2} \otimes \cdots \otimes O_{q,n}$ is a Pauli string, $U=U_1 \otimes U_2 \otimes \cdots \otimes U_n$ and $U_i=\left\{\begin{array}{ll}
R_y(-\pi/2) , & \text { if } O_{q,i}=X, \\
R_x(\pi/2), & \text { if } O_{q,i}=Y, \\
I, & \text { if } O_{q,i}=Z.
\end{array} \right.$}
      \label{circuit}
   \end{figure}

As seen from Eq~\eqref{Hc}, the problem Hamiltonian $H_P$ is given as a sum of Pauli strings, including $Z_i$ and $Z_iZ_j$ terms. Hence the unitary $V_p$  can be implemented efficiently as a quantum circuit using one of the time-independent quantum simulation algorithms \cite{tacchino2020quantum,berry2015simulating,cirstoiu2020variational,gibbs2024dynamical,magann2021digital}. In this work, we use the approach in \cite{tacchino2020quantum}, where the general form of the quantum circuit of $e^{-iO_{q} \Delta t}$ is shown in Figure~\ref{circuit}.  To enable efficient quantum circuit implementation for the Mixer operator $V_M(\theta_n)$ and calculation of the controller, we construct the mixer Hamiltonian $H_M$ as a sum of Pauli strings in the following form:
\begin{equation}
     H_M = \sum_{q=1}^{m_1} \hat{c}_q \hat{O}_q
\end{equation}
where $ \hat c_q $'s are real coefficients, $m_1$ is a polynomial function of the number of qubits, $\hat O_q$ is a Pauli string, i.e. $ \hat O_q = O_{q,1} \otimes O_{q,2} \otimes \dots \otimes O_{q,n} $ 
with each $ O_{q,d} \in \{I, X, Y, Z\} $. 

Such a design of the mixer Hamiltonian enables efficient expansion of the controller in the Pauli strings basis as follows:
\begin{equation}
    \theta_{n+1}= - \kappa \operatorname{Tr} (\left[H_M,Q_c \right] \sigma_n)= - \kappa \sum_{r=1}^{m_2} a_r \operatorname{Tr} ( R_r \sigma_n),
    \label{FALQON-IC_controller}
\end{equation}
where $R_r$ represents a Pauli string.

Consequently, the circuit parameters for the next layer are computed by estimating the expectation values of each Pauli string observable $R_r$ and applying Eq.~\eqref{FALQON-IC_controller}. Note that the number of Pauli strings $m_2$ depends on the Hamiltonians $H_P$ and $H_M$. Since both $m_0$ and $m_1$ are polynomial functions of the number of qubits, $m_2$ is also a polynomial function of the qubit count. The steps for implementing FALQON-C are outlined in Algorithm~1.

\algnewcommand\algorithmicInput{\textbf{Input:}}
\algnewcommand\algorithmicOutput{\textbf{Output:}}
\algnewcommand\Input{\item[\algorithmicInput]}
\algnewcommand\Output{\item[\algorithmicOutput]}
   
    \begin{algorithm} [H]
    \caption{FALQON-C \cite{abdul2024feedback}}\label{FALQON-C}
    \begin{algorithmic}[1]
    \Input{Problem Hamiltonian $H_P$, Mixer Hamiltonian $H_M$, Time step $\Delta t $, Maximum circuit depth $p$, Quantum circuit $V_0$ to prepare the initial state }
    \Output{The quantum circuit $V(\theta)$ for approximating the ground state of the operator $Q_c$ along with its parameters $\{\theta_n\}_{n=1}^p$}

    \State{Initialize the circuit parameter for the first layer $\theta_0=0$}
    \State{Design the operator $ Q_c$ using \eqref{Qc}}
    \State{\textbf{Repeat} at every step $n = 1, 2, 3, \dots, p-1$}
            \State {Prepare the initial state $\sigma_0=V_0 (\ketbra{0}^{\otimes n}) V_0^{\dagger} $}
            \State {Prepare the quantum state $\sigma_n = V_n \sigma_0  V_n^{\dagger}$, where $ V_n = \prod_{r=1}^{n} (V_M(\theta_r)   V_P )$ }
            \State {Calculate the circuit parameter of the next layer using $\theta_{n+1}= - \kappa \operatorname{Tr} (\left[H_M,Q\right] \sigma_n)$}
            \State {\textbf{Until $n=p$}}

    \end{algorithmic}
    \end{algorithm}

     \begin{figure}[H]
      \centering
      \includegraphics[width=1\linewidth]{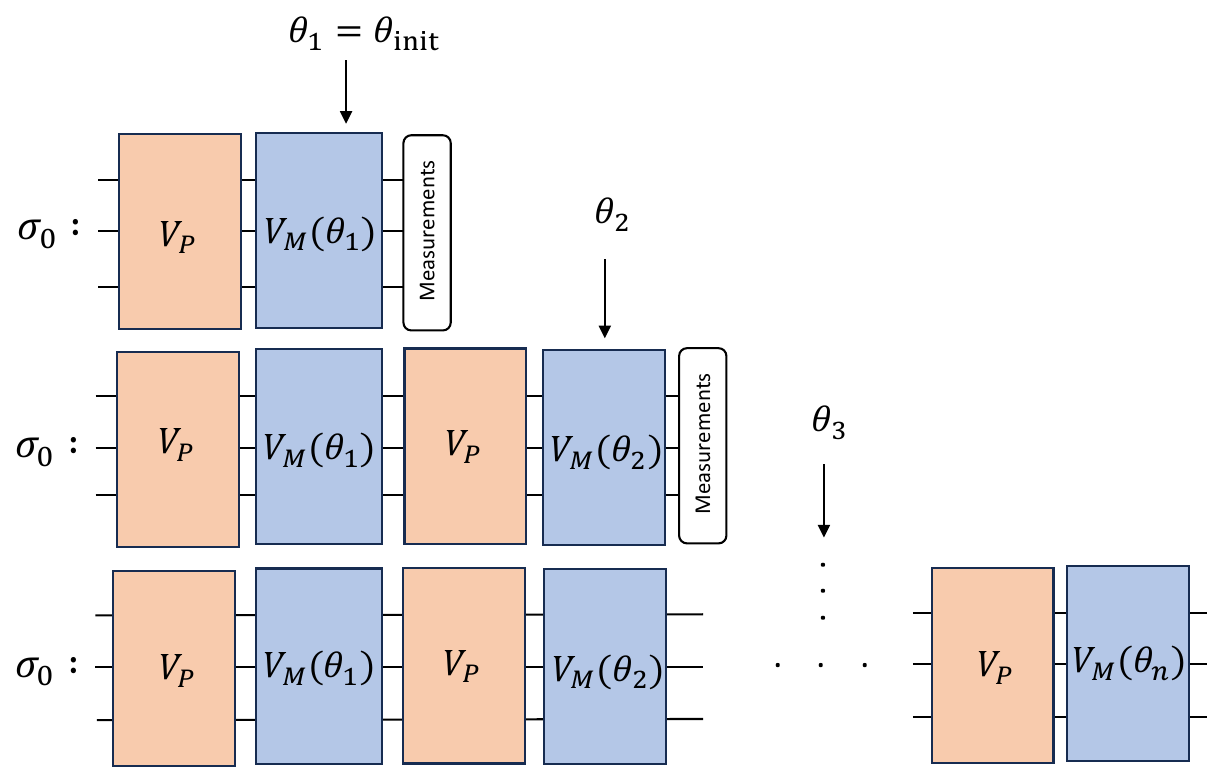}
      \caption{The figure illustrates the sequential process of executing FALQON-C. Initially, the algorithm begins with an initial guess for the first layer's parameter, $\theta_1=\theta_\text{init}$, which is used in implementing the first layer of the quantum circuit. This prepares the state $\sigma_1$ from an easy-to-prepare initial state $\sigma_0$ on the quantum computer. The parameter for the next layer, $\theta_2$, is then determined using Eq. \eqref{FALQON-IC_controller}, based on measurements obtained from the state $\sigma_1$. In each subsequent iteration, the quantum circuit grows incrementally by appending an additional layer consisting of $V_M(\theta_n)V_P$, where $\theta_n$ is determined based on the measurements from the previous layer using Eq. \eqref{FALQON-IC_controller}. This iterative procedure continues until the circuit reaches the predefined depth of $p - 1$ layers, at which point the execution is terminated.}%
      \label{Alg}
   \end{figure}

\subsection{Modifying Lyapunov Function to address the IC Constraints} \label{s}
We now introduce our first approach to tackle the IC constraints using FALOQN-C. In this approach, we use the conversion introduced in Theorem~1 to convert the IC constraints into an equivalent term, which will be used in the design of the operator $Q_c$. This term is then converted into a Hamiltonian $H^{(q)}_{\text{IC}}$ using the conversion $x_q \mapsto \frac{1}{2}(I - Z_q)$. The operator $Q_c$ is then given as:
\begin{equation}
    Q_c := H_P + \sum_{q=1}^{n_1} \gamma_qH^{(q)}_{\text{IC}}+\sum_{q=1}^{n_2} \beta_q H_C^{(q)},
    \label{Qc1}
\end{equation}
where the hyperparameters $\{\beta_q\}_{q=1, \dots, n_2}$ and $\{\gamma_q\}_{q=1, \dots, n_1}$ are chosen large enough to ensure that the ground state of the operator $Q_c$ encodes the optimal feasible solution of the QCBO problem.

We now show how to select the $\{\gamma_q\}$ hyperparameters. The approaches we propose here for choosing the hyperparameters $\{\gamma_q\}$ also apply to choosing the hyperparameters defined in Section~\ref{S3} for Theorem~1 where $F(\mathbf{x}_A^*) = e^\text f_{\min}$. To ensure that the smallest eigenvalue of $Q_c$ corresponds to an eigenvector that encodes the solution to the problem \eqref{QCBO}, it is enough to choose them such that $\gamma_q > \max{\{0,  e^\text f_{\min} - F(z^{(q)})\}}$. Since $e^\text f_{\min}$ is not known in advance, we propose three approaches to selecting the hyperparameters to satisfy this inequality.

One approach is to select the hyperparameters as $\{\gamma_q\} > e_g > \max{\{0,  e^\text f_{\min} - F(z^{(q)})\}}$. This choice will guarantee that the eigenvalues corresponding to IC constraints are shifted to be larger than the largest eigenvalue corresponding to a feasible outcome.

A more refined approach involves the following. Assume we can specify a feasible outcome $x^{(j)} \in \mathcal{F}$. Note that this could not be possible if we have inequality constraints besides the IC constraints. For the feasible outcome $x^{(j)} \in \mathcal{F}$, let $J(x^{(j)})$ denote the cost associated with this outcome, and let $J(z^{(q)})$ denote the cost associated with the IC constraint $z^{(q)}$. The hyperparameter $\gamma_q$ is then assigned based on the relationship between the cost of the specified feasible outcome, $J(x^{(j)})$, and $J(z^{(q)})$. If $J(x^{(j)}) < J(z^{(q)})$, the state associated with the IC constraint is not the ground state, meaning no further shifting is required, so we set $\gamma_q = 0$. If $J(x^{(j)}) > J(z^{(q)})$, the state associated with the IC constraint must be shifted to prevent it from being incorrectly assigned as the ground state, and we choose $\gamma_q$ such that $\gamma_q > J(x^{(j)}) - J(z^{(q)})$, ensuring that the eigenvalue corresponding to the IC constraint is raised above the cost of this feasible outcome. In the case where $J(x^{(j)}) = J(z^{(q)})$, it implies that there is a feasible outcome with the same cost as the IC constraint. In this situation, we also set $\gamma_q = 0$, as no shifting is needed. Suppose the algorithm converges to the state corresponding to the IC constraint, indicating that it has an equal value to the optimal solution. In that case, we can consider the specified feasible outcome as the solution instead. Such a choice results in lower values for the shifts. In the worst-case scenario, where our specified outcome corresponds to the maximum feasible eigenvalue, $e_{\max}$, and $J(z^{(q)}) = e^\text f_{\min}$, the resulting shift will be $e_{\max} - e^\text f_{\min}$. This is smaller than the shift required in the previous approach, which chose $\gamma_q > e_g$, where $e_g$ could be significantly larger than $e_{\max} - e^\text f_{\min}$.

A third approach introduces an iterative method for adjusting the parameters $\gamma_q$. Initially, $\gamma_q$ is set to a small trial value $\bar \gamma_q >0 $. If convergence to an infeasible outcome corresponding to the IC constraint is noticed, this suggests that $\bar \gamma_q < e^\text{f}_{\min} - J(z^{(q)})$. In this case, the parameter $\bar \gamma_q$ is doubled, and the process is repeated. This iterative procedure guarantees that $\bar \gamma_q > e^\text{f}_{\min} - J(z^{(q)})$ after $O\Big(\log_2\big(e^\text{f}_{\min} - J(z^{(q)})\big)\Big)$ iterations. Although this approach requires multiple evaluations, it enables the selection of smaller, more efficient values for the hyperparameters, ensuring sufficient penalization of infeasible outcomes while improving computational efficiency.

Appropriate choices for the hyperparameters $\{\beta_q\}$ and $\{\gamma_q\}$ can sometimes be inferred directly from the problem structure. While the approaches proposed in this work ensure that the operator $Q_c$ encodes the solution in its ground state, they tend to be conservative and may introduce unnecessarily large shifts. Optimizing the selection of these hyperparameters remains an open challenge. Simulation results indicate that excessively large hyperparameter values are unnecessary and can degrade the algorithm's performance. Instead, selecting smaller hyperparameter values that are sufficiently large to ensure that the ground state of $Q_c$ corresponds to the minimum eigenvalue improves the results.

\subsection{Modifying Lyapunov Function Using Deflation Approach} \label{ss.Df}
We now introduce the first approach for designing the operator $Q_c$ for FALQON-IC. In this approach, we utilize  Hotelling’s deflation technique to handle the IC constraints by appropriately shifting their corresponding eigenvalues (for more details on Hotelling’s deflation technique, refer to \ref{appC}). The equality and inequality constraints are handled as explained before. Consequently, we design the operator $Q_c$ such that its ground state encodes the solution to the QCBO problem given in \eqref{QCBO}. The resulting operator is given as follows:
\begin{equation}
    Q_c := H_P+ \sum_{q=1}^{n_1} \gamma_q \ketbra*{z^{(q)}}+\sum_{q=1}^{n_2} \beta_q H_C^{(q)} ,
    \label{Qc-DF}
\end{equation}
where $\ketbra*{z^{(q)}}$ are projectors onto the states associated with the IC constraints given in \eqref{ICs}, and $\{\beta_q\}_{q=1, \dots, n_2}$ and $\{\gamma_q\}_{q=1, \dots, n_1}$ are hyperparameters chosen large enough to ensure that the ground state of the operator $Q_c$ encodes the optimal feasible solution of the QCBO problem. In this case, we select the hyperparameters using the previously proposed approaches. From \eqref{Qc-DF}, it follows that the operator $Q_c$ shares the same eigenvectors as $H_P$, satisfying the commutation relation $[H_P, Q_c] = 0$. Recall that the set of infeasible outcomes due to the IC constraints is defined as $\mathcal{E}=\{x \in \{0,1\}^n : x = z^{(r)}, \; r\in \{1,2, \dots, n_1 \} \}$, and the set of feasible outcomes as $\mathcal{F} = \{x \in \{0,1\}^n : P^{(q)}(x)=0, \; q\in \{1,2, \dots, n_1 \} \}$. We define the set of infeasible outcomes due to the equality constraints as $\mathcal{V}=\{x \in \{0,1\}^n : P^{(q)}(x) \neq 0, \text{ for some } q\in \{1,2, \dots, n_2 \} \}$. The overall set of infeasible states is then  $\mathcal{F}^c=\mathcal{V} \cup \mathcal{E}$. Consequently, the eigenvalues of the operator $Q_C$ are expressed as:
\begin{equation}
\zeta_j(Q_c)= 
\begin{cases}
   e_j  & \text{for } j \in \mathcal{F} \\
   e_j+ \sum_q \beta_q P^{(q)}(j)  & \text{for } j \in \mathcal{V} \\
   e_j + \gamma_q. & \text{for } j \in \mathcal{E}
\end{cases}
\end{equation}

Note that by designing the operator $Q_c$ using the deflation approach, it is not possible to efficiently expand the observable $i[H_M,Q_c]$ in Eq. \eqref{controller1} in a Pauli strings basis. We, therefore, calculate the controller using different methods. The details on evaluating the controller for this case are given in \ref{appA.3}. 

\begin{remark}
    FALQON can be regarded as a special case of FALQON-IC when solving a QUBO problem with no constraints. In this case, the penalty Hamiltonian becomes trivial, and the operator $Q_c$ reduces to the problem Hamiltonian $H_P$, targeting the ground state of $H_P$. However, two approaches are possible when the problem is QCBO: solving it directly using FALQON-IC or converting the QCBO problem into a QUBO format and solving it with FALQON. In the numerical simulations section, we will demonstrate that, on a QCBO problem, FALQON-IC can save computational resources compared to FALQON as applied to the QUBO format of the problem.
\end{remark}

\subsection{Modifying Lyapunov Function Using Folded Spectrum Approach} \label{sec4}
This subsection introduces our second alternative approach for designing the operator $Q_c$ for FALQON-IC using the folded spectrum technique.
This method assumes that the outcomes corresponding to IC constraints represent the lowest-cost outcomes. Under this assumption, their costs can be evaluated, and the spectrum can be folded above the highest cost associated with these outcomes. However, if the IC constraints do not correspond to the lowest-cost outcomes, the folded spectrum technique cannot be applied, as folding the spectrum could shift the global optimal feasible solution, rendering the method invalid. Therefore, this approach only applies to problems where IC constraints are known beforehand to correspond to the lowest-cost outcomes. Given the assumption that the IC constraints correspond to the lowest-cost outcomes, the operator $Q_c$ will have its lowest eigenstates associated with these constraints. As a result, solving for the minimum energy feasible solution becomes equivalent to an excited-state calculation task for the Hamiltonian $H_P$. Consequently, as employed in this section for handling IC constraints, the FS method provides an alternative framework to the Feedback-based Quantum Algorithm for Excited States (FQAE) \cite{rahman2024feedbacke} for computing excited states.

\begin{figure}[H]
      \centering
      \includegraphics[width=1\linewidth]{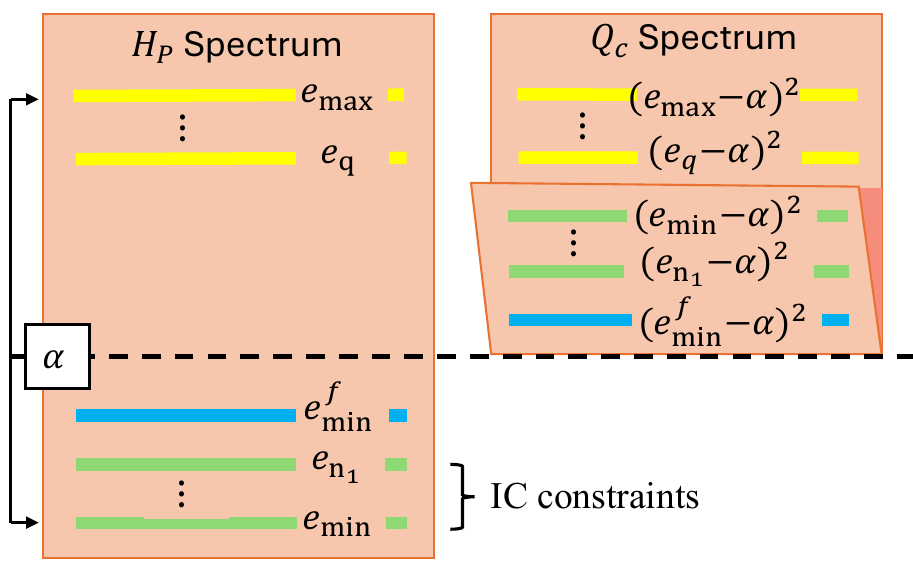}
      \caption{Illustration of the operator $Q_c$ design using the folded spectrum method. The operator $Q_c$ is constructed by folding the spectrum of $H_P$ around $\alpha$, which is designed to ensure that the eigenvector corresponding to the minimum feasible outcome becomes the ground state of $Q_c$, thereby encoding the solution to the QCBO problem.
      }%
      \label{FSS}
   \end{figure}

Similar to the previous method, the cost function is mapped to the problem Hamiltonian $H_P$, while the inequality constraints are transformed into equality constraints first and then, alongside the equality constraints, into the constraint Hamiltonians $H_C^{(q)}$. As the objective is to make the ground state of the operator $Q_C$ encode the solution to the QCBO, we fold the spectrum of $H_P$ around a hyperparameter $\alpha$ to get the following operator:
\begin{equation}
    Q_c := \Big( \big(H_P+\sum_{q=1}^{n_1} \beta_q H_C^{(q)} \big) -\alpha I\Big)^{2m},
    \label{FS}
\end{equation}
where we choose $\alpha$ to be closest to $e^\text f_{\min}$, i.e.  $\alpha \in(\frac{e^\text f_{\min} + e_{n_1}}{2},\frac{e^\text f_{\min} + \bar e}{2})$ and $\bar e$ represents the smallest eigenvalue in the spectrum of $H_P$ corresponding to a feasible outcome, that is larger than $e^\text f_{\min}$.
This choice of $\alpha$ guarantees that the operator $Q_c$ encodes the solution of the QCBO problem as its ground state. In this work, we choose $m=1$. From \eqref{FS}, it can be seen that the operator $Q_c$ and the problem Hamiltonian $H_P$ share the same set of eigenvectors and hence they commute ($[Q_c,H_P]=0$), while the eigenvalues of the operator $Q_c$ are reordered as follows:
\begin{equation}
\zeta_j(Q_c)= 
\begin{cases}
   (E_j-\alpha)^2  & \text{for } j \in \mathcal{F} \cup \mathcal{E} \\
   \big(E_j+ \sum_q \beta_q P^{(q)}(j)-\alpha\big)^2.  & \text{for } j \in \mathcal{V} 
\end{cases}
\end{equation}
The design approach for the operator $Q_c$ using the folded spectrum method is illustrated in Figure~\ref{FSS}. 
Since $e^\text f_{\min}$ and $\bar{e}$ are not known in advance, the hyperparameter $\alpha$ cannot be chosen directly. To address this, an iterative adjustment of $\alpha$ can be employed. The process begins by evaluating the cost $J(z^{(q)})$ for all $q \in \{1,2\dots,n_1\}$ to identify the IC constraint with the largest cost, denoted as $e_{n_1}$. Next, $\alpha$ is set as $\alpha = \bar{\alpha} + e_{n_1}$, where $\bar{\alpha} > 0$ is a small positive value ensuring $\alpha > e_{n_1}$. If the system converges to an IC constraint, the value of $\bar{\alpha}$ is doubled, and the process is repeated. Through this iterative refinement, after $O\Big(\log_2(\frac{e^\text f_{\min} + e_{n_1}}{2} - e_{n_1})\Big)$ iterations, a suitable $\alpha > \frac{e^\text f_{\min} + e_{n_1}}{2}$ can be determined.

Note that designing the operator $Q_c$ using FS as per Eq \eqref{FS} enables efficient expansion of the controller in the Pauli strings basis as follows:
\begin{equation}
    \theta_{n+1}= - \kappa \operatorname{Tr} (\left[H_M,Q_c\right] \sigma_n)= - \kappa \sum_{r=1}^{m_3} a_r \operatorname{Tr} ( R_r \sigma_n),
    \label{udis_exp}
\end{equation}
where $R_r$ represents a Pauli string. However, it will include more terms because of the square.

\section{Numerical Simulations}\label{S5}
In this section, we begin by introducing essential metrics used to evaluate the performance of the proposed algorithms. We then present a detailed analysis of the quantum resources required to implement their corresponding quantum circuits. Finally, we give numerical simulations for general QCBO problems to show the performance of the proposed algorithms.

\subsection{Performance Metrics}\label{SS5.1}
To assess the performance of the proposed algorithm, we utilize two key metrics: the approximation ratio and the success probability. The approximation ratio $r_a$ for a given state $\sigma$ is defined as the following expectation value \cite{herman2023constrained}:
\begin{equation}
r_a =\expval{\Pi_\mathcal{F}  \frac{{H_P} - e^\text f_{\max}}{e^\text f_{\min} - e^\text f_{\max}}\Pi_\mathcal{F}}_\sigma,
\end{equation}
where $\expval{\cdot}_\sigma := \operatorname{Tr}\big((\cdot) \sigma \big)$, $\Pi_\mathcal{F}$ is the projector onto the manifold of feasible states defined as $\Pi_\mathcal{F}:=\sum_{x \in \mathcal{F}} \ketbra{x}$. Here, $e^\text f_{\min}$ and $e^\text f_{\max}$ represent the smallest and the largest eigenvalue of $H_P$, respectively, corresponding to eigenvectors that encode feasible outcomes. 
The approximation ratio is defined so that the feasible states corresponding to the largest and smallest eigenvalues have values of 0 and 1, respectively. The success probability at a given state $\sigma$, denoted as $S_p(\sigma)$, is given by  
\begin{equation}  
S_p(\sigma) \equiv \sum_{x^* \in \mathcal{X}^*} \operatorname{Tr} \left(\ketbra{x^*} \sigma \right),  
\end{equation}  
where $\mathcal{X}^*$ represents the set of all bit strings corresponding to the problem's global optimal solution, which may be degenerate.

 \subsection{Quantum Resource Estimates}

Based on the Hamiltonian simulation method described in Subsection~\ref{ssFC}, we estimate the quantum resources, namely the number of qubits and the number of gates required to implement a single layer, for FALQON, FALQON-C, and FALQON-IC. Since the quantum circuit construction for FALQON-IC is identical for both the deflation and FS approaches, our analysis applies to both.

\textbf{Estimation of the number of qubits:} Consider a QCBO problem in the general form \eqref{QCBO}, where $n$ denotes the number of decision variables, $n_1$ represents the number of IC constraints, and $n_2$ is the total number of equality and inequality constraints. We first calculate the number of qubits to encode the problem. As described in Section 4, we start by converting the inequality constraints into equality constraints for all the proposed algorithms. This can be achieved by adding $n_{s_1}$ slack variables, where the exact number $n_{s_1}$ depends on the structure of the inequality constraints and the chosen conversion method (see \cite{montanez2024unbalanced,glover2022quantum} for further details) \footnote{For example, consider an inequality constraint of the form $c_G^T x \leq a_G$, where ${c_G}_q \in \mathbb{Z}$. This constraint can be converted into an equality by introducing $n_{s_1} = \left\lfloor \log_2\left( \max_x \left(a_G - c_G^T x\right) \right) + 1 \right\rfloor$ binary slack variables. The resulting equality constraint is given by $a_G-c_G^T x-\sum_{j=0}^{n_{s_1}-1} 2^j s_j=0$ \cite{glover2022quantum}.}. In addition, for FALQON and FALQON-C, we need to add $n_{s_2}= n_1 (n-2)$ slack variables to handle the IC constraints. As a result, the total qubit count for FALQON and FALQON-C becomes $n_t = n + n_{s_1} + n_1(n - 2)$. In contrast, FALQON-IC requires only $n_t = n + n_{s_1}$ qubits, since it incorporates IC constraints directly into the control framework without slack variables.

\begin{table*}[bt!]
\caption{Quantum resource estimates for implementing one layer of FALQON, FALQON-C, and FALQON-IC for solving QCBO problems in the general form of \eqref{QCBO}. The estimates assume the use of the standard Hamiltonian for the mixer $H_M = \sum_{q=1}^{n_t} X_q$, equal superposition initial state $\sigma_0 = \ketbra{+}^{\otimes n_t}$, and the Hamiltonian simulation method described in Subsection~\ref{ssFC}. Here, $n$ is the number of decision variables, $n_{s_1}$ the number of slack variables introduced for inequality constraints, and $n_1$ is the number of IC constraints. The numbers $\ell_1\leq n$ and $\ell_2 \leq n(n-1)/2$ represent the number of $Z_q$ and $Z_qZ_j$ terms in $H_P$ respectively. The numbers $\ell_3 = (n-2)(n+3)/2$ and $\ell_4 = (n-2)(n+3)/2$ represent the additional $Z_q$ and $Z_qZ_j$ terms introduced by the penalty Hamiltonians for IC constraints, while the numbers $\ell_5$ and $\ell_6$ represent the number of additional $Z_q$ and $Z_qZ_j$ terms introduced by the Hamiltonians from the inequality constraints, respectively (derivation of $l_1$, $l_2$, $l_3$ and $l_4$ is explained in the main text). Since $\ell_5$ and $\ell_6$ depend on the structure and number of inequality constraints, their closed-form expressions are not provided.} 

\label{tab:resources}
\begin{center}
\begin{tabular}{|l|c|c|c|}
\hline
\textbf{Resource Type} & \textbf{FALQON} & \textbf{FALQON-C} & \textbf{FALQON-IC} \\
\hline
\textbf{Number of Qubits} & 
$n + n_{s_1} + n_1(n{-}2)$ & 
$n + n_{s_1} + n_1(n{-}2)$ & 
$n + n_{s_1}$ \\
\hline
\textbf{Single-Qubit $R_x$ Gates} & 
$n + n_{s_1} + n_1(n{-}2)$ & 
$n + n_{s_1} + n_1(n{-}2)$ & 
$n + n_{s_1}$ \\
\hline
\textbf{Single-Qubit $R_z$ Gates} & 
\makecell[l]{
$\ell_1 + \ell_2 + \ell_3 + \ell_4 + \ell_5 + \ell_6 \leq$ $ \frac{n^2 + n}{2} + n_1(n^2+n-6) + \ell_5 + \ell_6$
} & 
\makecell[l]{
$\ell_1 + \ell_2 \leq \frac{n^2 + n}{2}$
} & 
\makecell[l]{
$\ell_1 + \ell_2 \leq \frac{n^2 + n}{2}$
} \\
\hline

\textbf{CNOT Gates} & 
\makecell[c]{
$2\ell_2 + 2\ell_4 + 2\ell_6 \leq n^2 - n + n_1(n^2 + n - 6)+2\ell_6$
} & 
$2\ell_2 \leq n(n{-}1)$ & 
$2\ell_2 \leq n(n{-}1)$ \\
\hline
\textbf{Hadamard Gates} & 
$n + n_{s_1} + n_1(n{-}2)$ & 
$n + n_{s_1} + n_1(n{-}2)$ & 
$n + n_{s_1}$ \\
\hline
\end{tabular}
\end{center}
\end{table*}

\textbf{Estimation of the number of gates:} We now calculate the number of gates required to implement the operators $V_M(\theta)$ and $V_P$. Assuming the mixer Hamiltonian is chosen as the standard mixer $H_M = \sum_{q=1}^{n_t} X_q$, implementing the operator $V_M(\theta)$ requires $n_t$ single-qubit $R_X$ rotation gates. Therefore, compared to FALQON-IC, both FALQON and FALQON-C incur an additional number $n_1(n - 2)$ of $R_X$ rotation gates due to the extra qubits introduced for handling IC constraints. For the operator $V_P$, the gate count depends on the structure of the problem and the constraints Hamiltonians. In all three algorithms, the base problem Hamiltonian $H_P$ is composed of $\ell_1$ single-qubit $Z_q$ terms and $\ell_2$ two-qubit $Z_q Z_j$ terms. Each $Z_q$ term is implemented using a single-qubit $R_Z$ rotation gate, while each $Z_q Z_j$ term requires two controlled-NOT (CNOT) gates and one $R_Z$ rotation gate. Thus, all three algorithms require $\ell_1 + \ell_2$ single-qubit $R_Z$ rotation gates and $2\ell_2$ CNOT gates from the $H_P$ Hamiltonian. We now calculate $\ell_1$ and $\ell_2$. By noting that $T_J$ is symmetric, Eq. \eqref{Hc} can be rewritten as 
\begin{align}
\label{Hcn}
H_P = \sum_{1\leq q<j\leq n}^{}\frac{1}{4} T_{J,q,j}Z_qZ_j - \sum_{q=1}^{n} \frac{1}{2}\big(c_{J,q}+ \sum_{j=1}^{n}T_{J,q,j}\big)Z_q.
\end{align}
Hence, we have a number of $\ell_1\leq n$ one-qubit $Z_q$ terms which is the number of nonzero terms in the sum $\sum_{q=1}^{n} \frac{1}{2}\big(c_{J,q}+ \sum_{j=1}^{n}T_{J,q,j}\big)$. These terms result in single-qubit $R_Z$ rotation gates. In addition, we have a number of $\ell_2 \leq \binom{n}{2} = n(n-1)/2$ two-qubit $Z_qZ_j$ terms which is the number of nonzero terms in the sum $\sum_{1\leq q<j\leq n}^{}\frac{1}{4} T_{J,q,j}$. This results in $\ell_2 \leq \frac{n(n-1)}{2}$ single-qubit $R_Z$ rotation gates and $2 \ell_2 \leq n(n-1)$ CNOT gates. In the case of FALQON, the constrained optimization problem is first transformed into the equivalent QUBO form. This results in the Hamiltonian $$\hat H_P = H_p + \sum_{q=1}^{n_1} \gamma_q H^{(q)}_{\text{IC}} + \sum_{q=1}^{n_2} \beta_q H_C^{(q)},$$ which introduces additional $Z_q$ and $Z_q Z_j$ terms derived from the penalty Hamiltonians for IC and inequality constraints. In the following, we estimate the additional gate cost due to a single IC constraint. The penalizing function $g(\cdot)$ can be expanded as follows:
\begin{align} \label{ge}
    g(x,s)&= 1 + v^T A h \nonumber \\
    &= 1+\sum_{q=1}^n v_q \cdot\left(-h_q+\sum_{k=q+1}^nh_k\right), \nonumber \\
    &= 1+\sum_{q=1}^{n-2} s_q \cdot\left(-h_q+\sum_{k=q+1}^nh_k\right)-h_{n-1}-h_n+h_nh_{n-1}.
\end{align}
Using the mapping $x_q \mapsto \frac{1}{2}(I - Z_q)$, the term $x_qx_j$ is mapped to the Hamiltonian $\frac{1}{4}(I-Z_q-Z_j+Z_qZ_j)$. In addition, the function $h_q(x)=x_q\oplus z_q$ can be equivalently expressed as $h_q(x)=z_q(1-x_q)+(1-z_q)x_q$. Using this representation, the term $h_qh_j$ can be expanded as $h_qh_j = z_qz_j + (z_q - 2z_qz_j)x_j +(z_j-2z_qz_j)x_q+(1-2z_q-2z_j+4z_qz_j)x_qx_j$ which is mapped to the Hamiltonain $\frac{1}{4}(I-(2z_q-1)Z_q-(2z_j-1)Z_j+(1-2z_q-2z_j+4z_qz_j)Z_qZ_j)$. Notably, the $Z_q$ and $Z_qZ_j$ terms that result from mapping the terms $-h_{n-1}-h_n+h_nh_{n-1}$ into a Hamiltonian will be included in the $\ell_1$ and $\ell_2$ terms and will not add additional gates. Hence, there will only be additional terms coming from mapping the term $\sum_{q=1}^{n-2} s_q \cdot\left(-h_q+\sum_{k=q+1}^nh_k\right)$ into a Hamiltonian. This will give a number $\sum_{q=1}^{n-2}(n-q+1) = \frac{(n+3)(n-2)}{2}$ of terms of the form $\pm s_qh_j $ which is equal to $\pm( z_js_q +(1-2z_j)s_qx_j)$. This term is mapped to the Hamiltonian $\pm \frac{1}{4}(I-Z_{q+n}-(1-2z_j)Z_j+(1-2z_j)Z_jZ_{q+n})$. This Hamiltonian will add $\ell_3 = \frac{(n+3)(n-2)}{2}$ new single-qubit $Z_{q+n}$ terms while the terms $Z_j$ will be added to $\ell_1$ and hence will not add new terms. Therefore, these terms will add a number $\ell_3 = \frac{(n+3)(n-2)}{2}$ of single $R_z$ gates. Moreover, we will have a number $\ell_4 = \frac{(n+3)(n-2)}{2}$ of $Z_jZ_{q+n}$ terms which will add $\frac{(n+3)(n-2)}{2}$ single-qubit $R_z$ gates and $(n+3)(n-2)$ CNOT gates. For a number $n_1$ of IC constraints, we will have $\ell_3 = \frac{n_1(n+3)(n-2)}{2}$ and $\ell_4 = \frac{n_1(n+3)(n-2)}{2}$. In contrast to FALQON, FALQON-C and FALQON-IC directly incorporate constraints into the control framework without modifying the generator Hamiltonian, thus avoiding these additional gate costs that come from mapping the penalizing function $g(\cdot)$. Note that inequality constraints, once converted to equality constraints and mapped into constraint Hamiltonians, may introduce additional $Z_q$ and $Z_q Z_j$ terms, denoted by $\ell_5$ and $\ell_6$, respectively. The exact values of these terms depend on the specific structure and number of the inequality constraints and are therefore not explicitly quantified here.

Finally, assuming the initial state is the equal superposition state $\sigma_0 = \ketbra{+}^{\otimes n_t}$, FALQON and FALQON-C require $n_1(n - 2)$ additional Hadamard gates to prepare the initial state. Table~\ref{tab:resources} provides a summary of these resource estimates. As shown, FALQON-IC can significantly reduce the qubit count and the number of quantum gates compared to FALQON and FALQON-C. This reduction highlights its advantage in terms of scalability and implementation on quantum hardware.

 \subsection{Illustrative example}
Consider the following instance of the SVP formulated as a QCBO problem as given in \cite{lv2022using}:
	\begin{align} \label{SVP} 
		&\min _{x \in \{0,1\}^{3}} S(x)= x_1+2x_2+5x_3+2x_2x_3 \nonumber\\
		&\text{s.t.} \quad x \neq z    
   \end{align}
where $z = [0,0,0]$. By transforming $S(x)$ into a problem Hamiltonian we get $H_P=4.5I-0.5Z_1-1.5Z_2-3Z_3+0.5Z_2Z_3=\text{diag}(0,5,2,9,1,6,3,10)$, or equivalently,

\begin{equation*}
 H_p = \begin{bmatrix}
0 & 0 & 0 & 0 & 0 & 0 & 0 & 0\\
0 & 5 & 0 & 0 & 0 & 0 & 0 & 0 \\
0 & 0 & 2 & 0 & 0 & 0 & 0 & 0 \\
0 & 0 & 0 & 9 & 0 & 0 & 0 & 0 \\
0 & 0 & 0 & 0 & 1 & 0 & 0 & 0 \\
0 & 0 & 0 & 0 & 0 & 6 & 0 & 0 \\
0 & 0 & 0 & 0 & 0 & 0 & 3 & 0 \\
0 & 0 & 0 & 0 & 0 & 0 & 0 & 10 
\end{bmatrix}.
\end{equation*}
It can be seen that the eigenvalue $0$, corresponding to the eigenvector that encodes the infeasible outcome $[0,0,0]$, is the lowest eigenvalue of $H_P$. Hence, the ground state of $H_P$ is an infeasible state, corresponding to the IC constraint $x \neq 0$. Notably, this corresponds to the zero vector, representing the trivial solution to the SVP (see \cite{albrecht2023variational} for details). In the following, we solve this problem using FALQON-IC. For comparison, we also apply FALQON and FALQON-C, utilizing the conversion method introduced in Theorem~1.

\begin{figure}[H]
      \centering
      \includegraphics[width=1\linewidth]{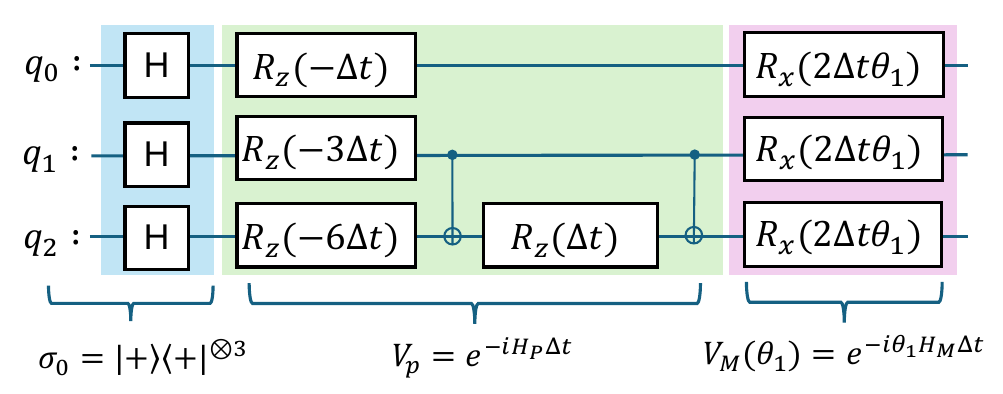}
       \caption{Quantum circuit for the first layer of FALQON-IC implementation.}
      \label{FALQON-IC_circuit}
   \end{figure}

\begin{figure}[H]
      \centering
      \includegraphics[width=1\linewidth]{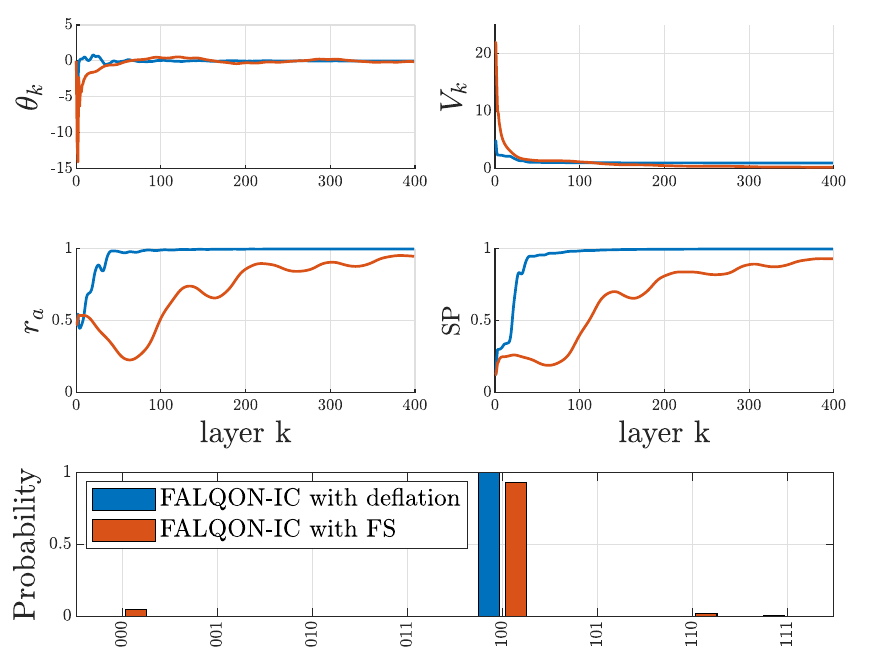}
       \caption{Comparison between the simulation results for running FALQON-IC with deflation approach and folded spectrum approach to solve the QCBO problem \eqref{SVP}. The layer index $k$ is plotted versus the controller $\theta_k$, the Lyapunov function $V_k$, the approximation ratio $r_a$, the success probability $SP$ and the histogram of the final state.}
      \label{compare_FS_FD}
   \end{figure}

To apply FALQON-IC, we need to design the operator $Q_c$. We first design it using the deflation approach as defined in \eqref{Qc-DF}, resulting in $Q_c=H_P+3 \ketbra{000}=\text{diag}(3,5,2,9,1,6,3,10)$, where we choose $\gamma =3$. It can be seen that this operator has the state $\ket{100}$ as its ground state, which encodes the solution to Problem~\eqref{SVP} since it corresponds to the lowest eigenvalue $1$. In addition, we design the operator using the FS approach as defined in \eqref{FS}. In this case, we obtain $Q_c = (H_P - 1.3I)^2 =(3.2I-0.5Z_1-1.5Z_2-3Z_3+0.5Z_2Z_3)^2=21.99I-3.2Z_1-12.6Z_2-20.7Z_3+1.5Z_1Z_2+3Z_1Z_3+12.2Z_2Z_3-0.5Z_1Z_2Z_3 = \text{diag}(1.69,13.69,0.49,59.29,0.09,22.09,2.89,75.69)$, where we choose $\alpha=1.3$. Similarly, in this case, the operator $Q_c$ encodes the state $\ket{100}$ as its ground state. We simulate FALQON-IC for both approaches using the statevector simulator. We design the mixer Hamiltonian as $H_M = \sum_{q=1}^3X_q$, and we choose the equal superposition as the initial state $\sigma_0=\ketbra{+}^{\otimes 3}$. We set the controller gain $\kappa=1$ and the initial value for the controller $\theta_1=0$. We set $\Delta t  = 0.1$ for the deflation approach and $\Delta t = 0.03$ for the FS approach. The quantum circuit for implementing the first layer of FALQON-IC is shown in Figure~\ref{FALQON-IC_circuit}, while the simulation results are shown in Figure~\ref{compare_FS_FD}. 

As can be followed from Figure~\ref{FALQON-IC_circuit}, the quantum circuits for implementing FALQON-IC using both the deflation and FS approaches are identical, requiring no additional qubits. As shown in Figure~\ref{compare_FS_FD}, the deflation approach achieves faster convergence than the FS approach.

We now solve the problem using FALQON and FALQON-C. We start by converting the IC constraint into a penalizing term using \eqref{g}. We use this term to design the observable $Q_c$ as defined in \eqref{Qc1} for FALQON-C. Additionally, to solve the problem using FALQON, we use this term to convert Problem \eqref{SVP} into an equivalent QUBO problem as per Theorem~1.

As shown in Section~\ref{sec4}, we need to add one slack variable $s_1$ to convert the IC constraint. From Eq. \eqref{h(x)}, we have $h(x)=[x_1,x_2,x_3]$ and $v(x,s)=[s_1,1-x_3,1]$. Using Eq. \eqref{g}, we get  $g(x,s)=1+s_1(-x_1+x_2+x_3)+(1-x_3)(-x_2+x_3)-x_3=1-s_1x_1+s_1x_2+s_1x_3-x_2-x_3+x_2x_3$ where we used $x_q^2=x_q$ for binary decision variables. Let $y=[y_1,y_2,y_3,y_4]:=[x_1,x_2,x_3,s_1]$ and $ \bar S(y):=y_1+2y_2+5y_3+2y_2y_3$. To apply FALQON-C, we construct the operator $Q_c$ by first mapping the cost function $\bar S(y)$ into a problem Hamiltonian to get $\bar H_P=4.5I-0.5Z_1-1.5Z_2-3Z_3+0.5Z_2Z_3=\text{diag}(0,0,5,5,2,2,9,9,1,1,6,6,3,3,10,10)$. Note that $\bar S$ and $\bar H_P$ are straightforward extensions of $S$ and $H$, respectively, to a larger space obtained by adding the slack variables. We then map $g(y)=1-y_2-y_3+y_2y_3-y_1y_4+y_2y_4+y_3y_4$ into a Hamiltonian, to get the Hamiltonian $H_\text{IC}=0.5I+0.25Z_1-0.25Z_4-0.25Z_1Z_4+0.25Z_2Z_3+0.25Z_2Z_4+0.25Z_3Z_4 = \text{diag}(1,1,0,1,0,1,0,2,1,0,0,0,0,0,0,1)$. Subsequently, we design the operator $Q_c$ using Eq. \eqref{Qc1} as $Q_c = \bar H_p + 3 H_{\text{IC}} = \text{diag}(3,3,5,8,2,5,9,15,4,1,6,6,3,3,10,13)$. It is seen that the ground state of the operator $Q_c$, which is the state $\ket{1001}$ encodes the optimal feasible solution to the problem $[1,0,0]$ in the first three qubits since the solution is the projection of $y=(x,s)$ on the first three qubits. The first three qubits encode the solution, while the last qubit represents the slack variable, which does not contribute to the cost function.

\begin{figure}[H]
      \centering
      \includegraphics[width=1\linewidth]{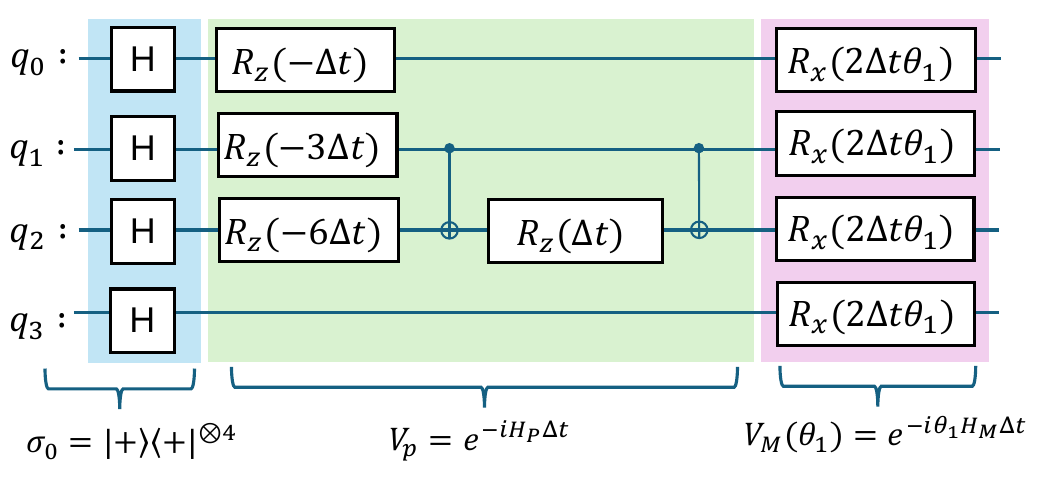}
       \caption{Quantum circuit for the first layer of FALQON-C implementation.}
      \label{FALQON-C_circuit}
   \end{figure}

The first layer of the quantum circuit of FALQON-C is shown in Figure~\ref{FALQON-C_circuit}.  

\begin{figure}[H]
      \centering
      \includegraphics[width=1\linewidth]{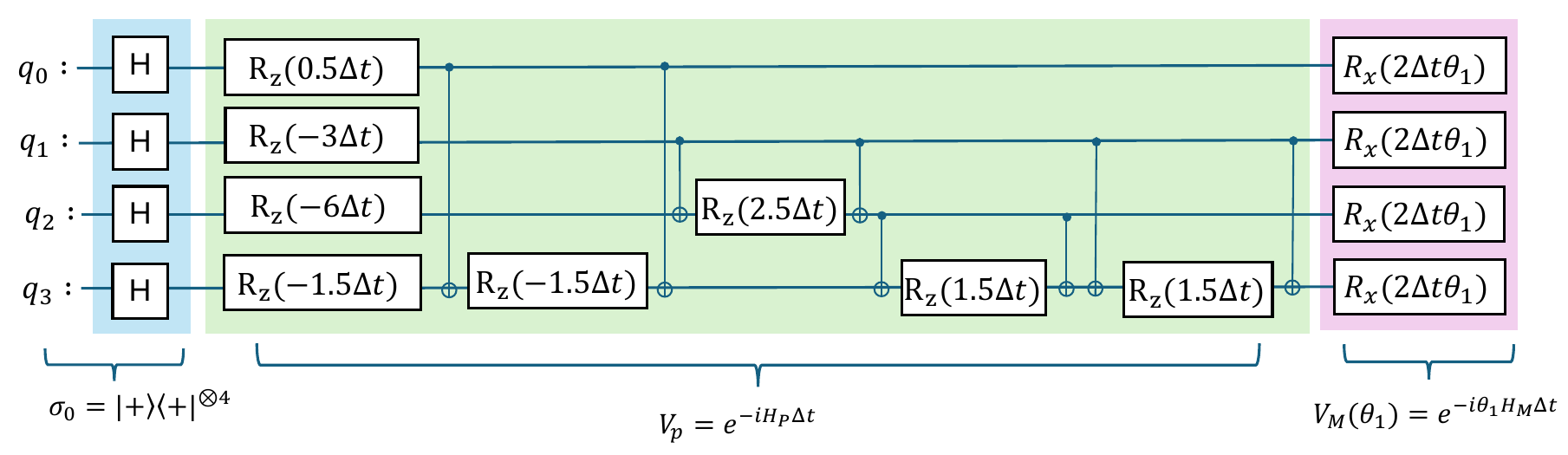}
       \caption{Quantum circuit for the first layer of FALQON implementation.}
      \label{FALQON_circuit}
   \end{figure}

\begin{figure}[H]
      \centering
      \includegraphics[width=1\linewidth]{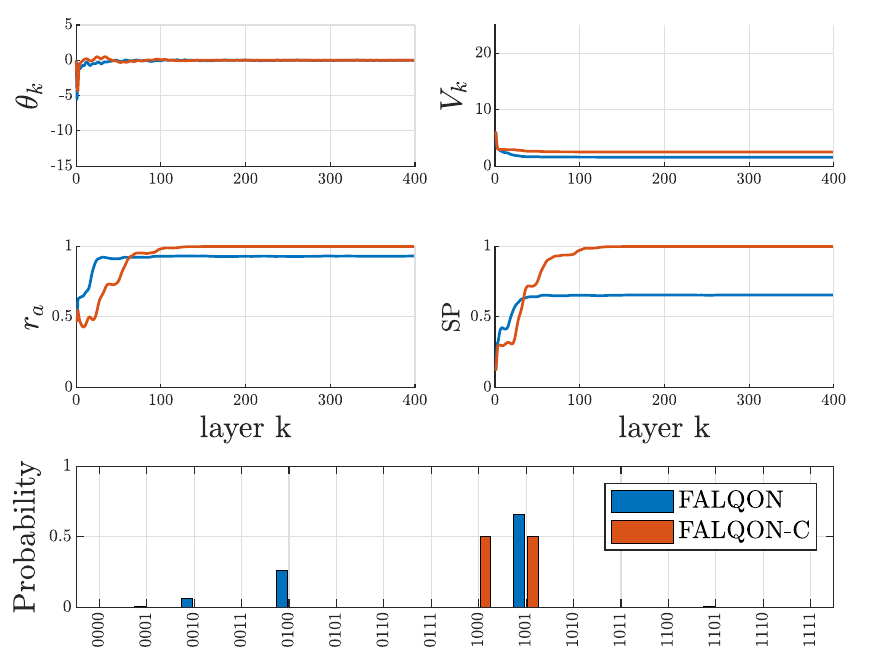}
      \caption{Comparison between the simulation results for running FALQON-C and FALQON to solve the QCBO problem \eqref{SVP}. The layer index $k$ is plotted versus the controller $\theta_k$, the Lyapunov function $V_k$, the approximation ratio $r_a$, and the success probability $SP$ and the histogram of the final state.}
      \label{compare_F_FC}
   \end{figure}

Finally, to apply FALQON, we use Theorem~1 to convert the problem into the following equivalent QUBO problem:
\begin{align} \label{SVPQ} 
		\min _{y \in \{0,1\}^{4}} L(y) & :=  \bar S(y)+ \gamma g(y) \nonumber\\
		& \; \; =   y_1 + 2y_2 +5_3 +2y_2y_3 \nonumber\\
        & \quad \;\;+3 (1-y_2-y_3+y_2y_3-y_1y_4+y_2y_4+y_3y_4).
\end{align}
Consequently, we map $L(y)$ into the Hamiltonian $\hat H_P = 6I+0.25Z_1-1.5Z_2-3Z_3-0.75Z_4- 0.75Z_1Z_4+1.25Z_2Z_3+0.75Z_2Z_4+0.75Z_3Z_4 = \text{diag}(3,3,5,8,2,5,9,15,4,1,6,6,3,3,10,13)$. We apply FALQON to find the ground state of this Hamiltonian, which encodes the solution to the problem. The first layer of the quantum circuit for FALQON is shown in Figure~\ref{FALQON_circuit}. We simulate FALQON and FALQON-C where we set the controller gain $\kappa=1$, the initial value for the controller $\theta_1=0$, and $\Delta t = 0.08$. Simulation results for FALQON and FALQON-C are given in Figure~\ref{compare_F_FC}. 

Comparing the quantum circuits of FALQON and FALQON-C in Figures \ref{FALQON-C_circuit}-\ref{FALQON_circuit}, it is seen that $4$ qubits are needed for implementing FALQON and FALQON-C. It is also seen that the quantum circuit of the operator $V_P$ is deeper for FALQON compared to FALQON-C. 
Figure~\ref{compare_F_FC} shows that FALQON-C when applied to the problem \eqref{SVP}, achieves a higher approximation ratio and success probability than FALQON. It is also noted from the histogram that for FALQON-C, the state converges to a superposition between the states $\ket{1000}$ and $\ket{1001}$. These two states encode the solution to the problem in their first three qubits since the last qubit represents the slack variable.

\begin{figure}[H]
      \centering
      \includegraphics[width=1\linewidth]{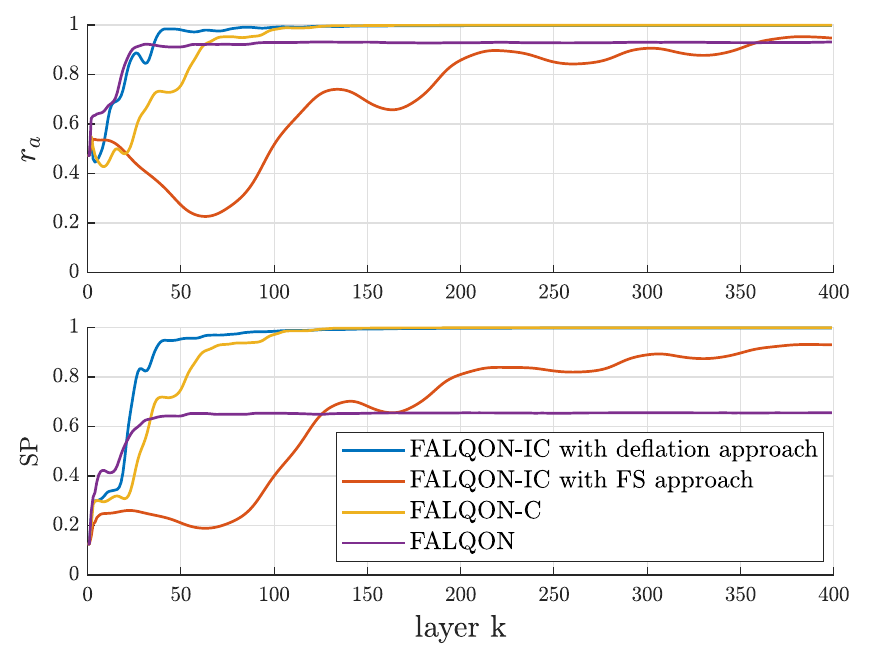}
       \caption{Comparison between the simulation results for running FALQON, FALQON-C and FALQON-IC to solve the QCBO problem \eqref{SVP}. The layer index $k$ is plotted versus the approximation ratio $r_a$ and the success probability $SP$.}
      \label{compare_all}
   \end{figure}

In Figure~\ref{compare_all}, we present a comparative analysis of all approaches in terms of approximation ratio and success probability. The results indicate that when applied to the QCBO problem \eqref{SVP}, FALQON-IC outperforms the other methods, achieving superior performance. In addition, as can be followed from Figures \ref{FALQON-IC_circuit}, \ref{FALQON-C_circuit} and \ref{FALQON_circuit}, it is seen that FALQON-IC needs less number of qubits, three in this case, compared to FALQON and FALQON-C which require four qubits. In addition, FALQON-IC has the shallowest quantum circuit, while FALQON has the deepest quantum circuit. 

\subsection{Scalability and Performance on Random QCBO Instances}
\begin{figure*}[b]
      \centering
      \includegraphics[width=0.9\linewidth]{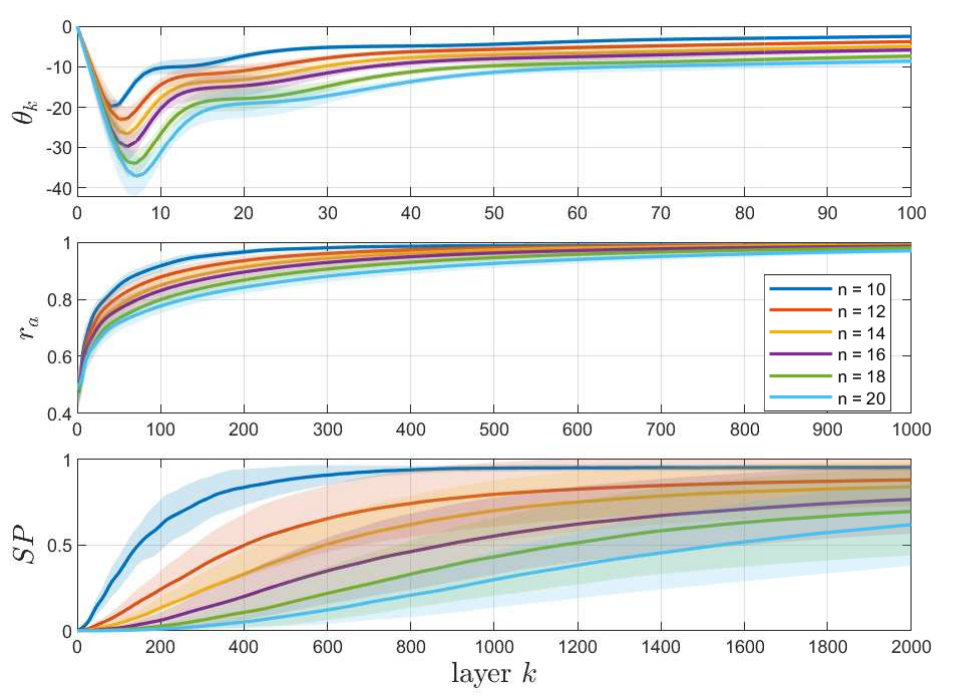}
       \caption{Simulation results of applying FALQON-IC to 50 randomly generated QCBO problems and for different problem sizes $n \in \{10, 12, \dots,20 \}$. The layer index $k$ is plotted versus the mean trajectory (solid line) and the corresponding standard deviation (shaded area) of the controller $\theta_k$, the approximation ratio $r_a$ and the success probability $SP$.}
      \label{compare_all_evos}
   \end{figure*}

To assess the scalability of FALQON-IC, we evaluate its performance on randomly generated instances of the QCBO problem~\eqref{QCBO}, which captures a broad class of combinatorial optimization problems widely studied in the quantum optimization literature~\cite{albrecht2023variational,glover2022quantum}. We generate 50 random instances for different problem sizes $n \in \{10,12, \dots, 20 \}$. To generate each instance, the entries of the cost matrix $T_J$, the linear term vector $c_J$ and the scalar term $a_J$ are independently sampled from a uniform distribution over the interval $[-5, 5]$, i.e., $T_{J_{q,j}} \sim \mathcal{U}(-5,5)$, $c_{J_q} \sim \mathcal{U}(-5,5)$ and $a_J \sim \mathcal{U}(-5,5)$. We also, without loss of generality, enforce symmetry in $T_J$, by replacing $T_{J_{q,j}}$ with $(T_{J_{q,j}}+T_{J_{j,q}})/2$ for all $q$ and $j$ \cite{glover2022quantum}. Additionally, we generate one IC constraint for each instance by uniformly sampling a binary vector $z \in \{0,1\}^n$, which defines the prohibited configuration $x \neq z$. We use the deflation approach in FALQON-IC for these simulations, as the FS approach only applies when it is known in advance that the IC constraint corresponds to the lowest-energy outcomes. For all simulations, we fix the controller gain at $\kappa = 1$ and the deflation hyperparameter at $\gamma = 8$. The time step $\Delta t$ is tuned to the largest value that ensures the condition $L(\sigma_{n+1}) - L(\sigma_n) \leq 0$ is satisfied across all randomly generated instances. Each simulation is run for a circuit depth of 2000 layers, and the results are presented in Figure~\ref{compare_all_evos}.

Figure~\ref{compare_all_evos} presents the evolution of the controller $\theta_k$, the approximation ratio $r_a$, and the success probability $SP$ as functions of the layer index $k$, for problem sizes $n \in \{10, 12, \dots , 20\}$. The solid curves denote the mean trajectories over 50 random QCBO instances, while the shaded areas represent their corresponding standard deviation. The results demonstrate monotonic convergence in both $r_a$ and $SP$ as the layer index increases.

\begin{figure}[H]
      \centering
      \includegraphics[width=1\linewidth]{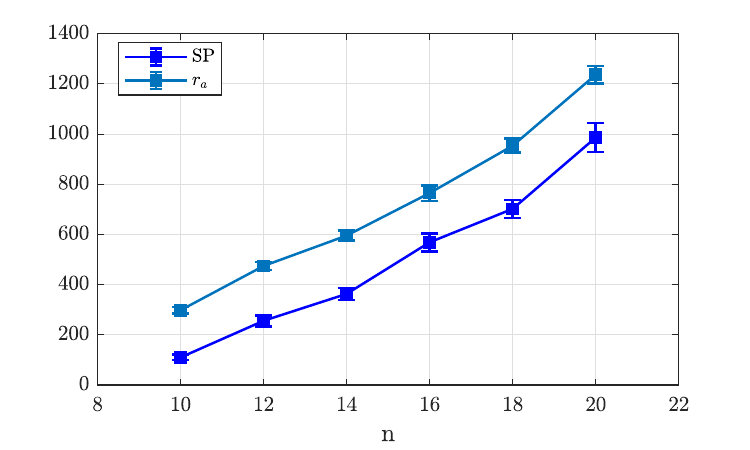}
       \caption{The mean circuit layers required to achieve an approximation ratio of $r_a=0.98$ and a success probability of $SP=0.25$ is plotted versus the problem size $n$. Results are averaged over the $50$ randomly generated QCBO instances for each problem size.}
      \label{sp_ra}
   \end{figure}

To assess the scalability of FALQON-IC, we further evaluate the average number of layers required to reach target performance thresholds. In Figure~\ref{sp_ra}, we plot the mean number of circuit layers needed to achieve an approximation ratio of $r_a = 0.98$ and a success probability of $SP = 0.25$, for each problem size. The results suggest that the proposed FALQON-IC algorithm exhibits favorable runtime scalability, with the required circuit depth scaling approximately linearly for general QCBO problems of the form~\eqref{QCBO}.

\begin{figure}[H]
      \centering
      \includegraphics[width=1\linewidth]{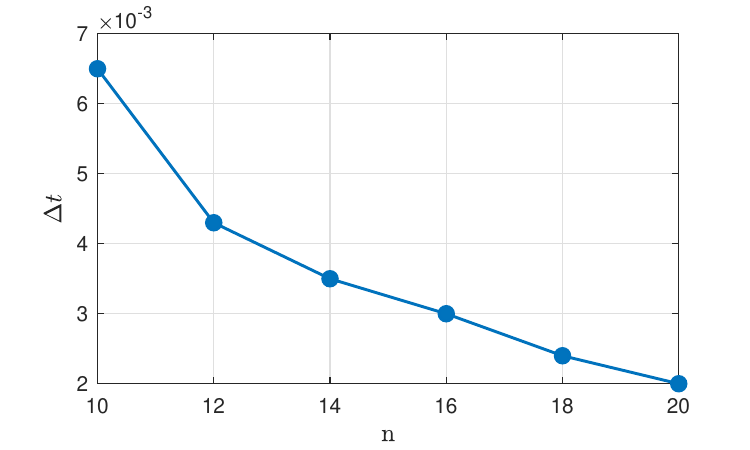}
       \caption{The time step $\Delta t$ for each of the problem sizes is plotted versus the problem size $n$.}
      \label{dt}
   \end{figure}

Figure~\ref{dt} displays the time step $\Delta t$ used for each problem size, revealing a decreasing trend that also scales roughly linearly with $n$. Although larger $\Delta t$ values can accelerate convergence, they must be chosen carefully to preserve Lyapunov stability. Larger values of $\Delta t$ can be utilized by incorporating higher-order terms into the feedback law design in \eqref{DV}. This approach can accelerate convergence and enhance performance, as discussed in \cite{arai2025scalable}. The work \cite{arai2025scalable} demonstrates that including higher-order terms in the control law allows for larger time steps without compromising convergence guarantees, providing a promising avenue for future developments.

Together, these results confirm that FALQON-IC offers robust and scalable performance for solving general QCBO problems with IC constraints, maintaining efficient convergence and reasonable quantum resource requirements even as the problem size increases.

\begin{figure}[H]
      \centering
      \includegraphics[width=1\linewidth]{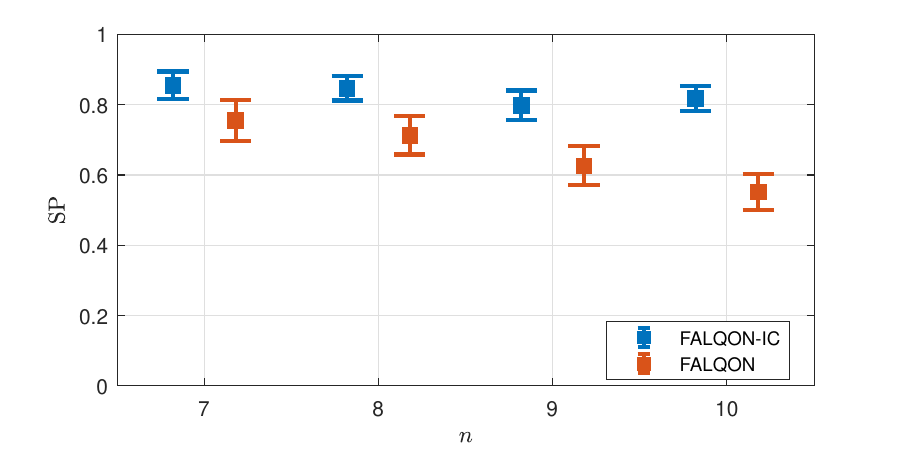}
       \caption{The mean of the success probabilities versus the problem size $n$. The error bars represent the standard error of the mean of these success probabilities. Results are averaged over the $50$ randomly generated QCBO instances for each problem size.}
      \label{sp_comp}
   \end{figure}

To evaluate the scalability and performance of FALQON-IC relative to FALQON, we simulate $50$ randomly generated QCBO instances, each with a single IC constraint, for each problem size $n \in \{7, 8, 9, 10\}$. In order to apply FALQON, Theorem 1 is used to transform each QCBO instance into an equivalent QUBO problem. This transformation requires the addition of $n - 2$ slack variables, increasing the total number of qubits, whereas FALQON-IC encodes the IC constraint directly and avoids introducing any slack variables. For both algorithms, we fix the controller gain at $\kappa = 1$ and the hyperparameter at $\gamma = 8$. The time step $\Delta t$ is tuned individually for each problem size to ensure convergence across all the instances. Specifically, for FALQON-IC, we use time steps $\Delta t = [0.008,\ 0.0048,\ 0.0048,\ 0.004]$ corresponding to problem sizes $n = [7,\ 8,\ 9,\ 10]$, respectively, while FALQON uses $\Delta t = [0.0058,\ 0.0035,\ 0.003,\ 0.0025]$ for the same sizes. Each algorithm is simulated for 1000 layers. Figure~\ref{sp_comp} presents the mean success probabilities as a function of problem size $n$, with error bars indicating the standard error of the mean over the $50$ problem instances. The results show that FALQON-IC outperforms FALQON, particularly as the problem size increases. This difference is primarily due to the overhead introduced by slack variables in FALQON, which increases the number of qubits and slows down convergence. This effect is further reflected in the smaller time steps $\Delta t$ required for FALQON to maintain convergence. Moreover, the advantage of FALQON-IC becomes more evident with an increasing number of IC constraints, as each additional constraint introduces $n - 2$ slack variables in FALQON, further increasing the size of the equivalent problem. These results show that FALQON-IC can significantly enhance the performance as well as the required quantum resources compared to FALQON for QCBO with IC constraints.

\section{Conclusion and Future Work} \label{S6}
In this paper, we extended the capabilities of feedback-based quantum algorithms to tackle a broader class of optimization problems, specifically, quadratic constrained binary optimization problems with IC constraints. We introduced Theorem~1 to convert the QCBO problem with IC constraints into an equivalent QUBO problem. This conversion enables using FALQON to solve the problem. It is also significant for quantum algorithms that require the problem to be in a QUBO formulation, such as quantum annealing and QAOA. Beyond this transformation, we introduced FALQON-IC, an algorithm to directly tackle IC constraints by designing the observable within the Lyapunov function. This method constructs a new operator that encodes the optimal feasible solution as its ground state, leveraging deflation techniques and the folded spectrum method to directly address IC constraints. A key advantage of this approach is its ability to eliminate the need for slack variables, reducing the number of required qubits and the depth of the quantum circuit, making it significantly more practical for near-term quantum hardware. However, this comes at the cost of increased computational complexity in determining the circuit parameters. Specifically, the deflation approach requires evaluating the gradient of the Lyapunov function, while the FS approach necessitates estimating additional terms, adding to the computational overhead.

Despite their potential, FQAs face several limitations. They rely on stringent convergence conditions that are often challenging to meet in practice and can become trapped in suboptimal solutions of the cost function. Additionally, FQAs produce deeper quantum circuits than VQAs with similar performance, which is a significant bottleneck, especially on NISQ devices. Several future directions exist to improve the performance of FQAs. As seen in Section~2, convergence is guaranteed for almost all initial conditions, and it is essential to employ a warm starting technique for FQAs. Another important direction is to investigate the efficient design of mixer Hamiltonians as explored in \cite{malla2024feedback}. Additionally, exploring alternative Hamiltonian simulation methods such as the linear combination of unitaries \cite{chakraborty2024implementing}, Taylor series expansion \cite{berry2015simulating}, and quantum signal processing \cite{low2017optimal} could help reduce circuit depth and improve computational efficiency. Lastly, relaxing the stringent conditions imposed on the Hamiltonians (such as those in Appendix A of \cite{magann2022feedback}) could enhance the performance and expand the class of problems that can be effectively solved while maintaining convergence guarantees.

Although this work focuses on feedback-based quantum algorithms, the proposed approaches can be directly adapted to VQAs to enhance their capability in solving constrained optimization problems. In particular, the deflation and the folded spectrum approaches for the design of the observables can be integrated into algorithms like QAOA and variational quantum eigensolver to address IC constraints more effectively. By leveraging these strategies, VQAs can encode feasible solutions directly into their cost functions. This adaptability highlights the broader impact of our work, offering new directions for constraint handling in both feedback-based and variational quantum algorithms.

Overall, our proposed FALQON-IC framework broadens the scope of feedback-based quantum algorithms, enabling them to tackle more complex constrained optimization problems. This work demonstrates the potential of control theory to drive the development of efficient quantum algorithms, broadening the applicability of FQAs to a broader range of constrained optimization challenges and laying a foundation for future advancements in quantum optimization and control.

\section*{Acknowledgements} \label{S7}
This work was supported by Independent Research Fund Denmark (DFF), project number 0136-00204B.

\appendix
\section{Controller Design and $\Delta t$ Parameter Selection} \label{appA}

\subsection{Handling Multiple Controllers} \label{appA.1}
The analysis in Section~\ref{S2} can be extended for the case of multiple controllers (circuit parameters per layer) as follows (for further details, see Subsection 2-C of \cite{magann2022lyapunov}).

In this case, the mixer becomes $V_M(\theta_n)=e^{-\sum_{r=1}^{r_m}i\theta_n^{(r)} H_m^{(r)}\Delta t }$, where $r_m$ is the total number of controllers. Consequently, the controller is modified to $\theta_n^{(r)}=-\kappa_r \Lambda \bigg(\operatorname{Tr} \big(i\left[H_M^{(r)},Q\right] \sigma_n\big) \bigg)$.

\subsection{Selection of the $\Delta t$ parameter} \label{appA.2}

By following similar analysis to Section 4-A of \cite{magann2022lyapunov}, the following modified bound for $\Delta t$ is obtained:

\begin{equation}
    |\Delta t|<\frac{\left| \expval{  [H_M, Q] }_{\sigma_n}  \right|}{2\left(2 ||H_M|| \cdot||H_P||+\left|\expval{  [H_M, Q] }_{\sigma_n}
 \right|\right)\left(||H_P||+||H_M|| \left| \theta_n \right|\right)}.
\end{equation}
Selecting $\Delta t$ within this bound ensures that the Lyapunov function is non-increasing, satisfying the condition $L(\sigma_{n+1}) - L(\sigma_n) \leq 0$.

\subsection{Controller Calculation for FALQON-IC with Deflation Approach} \label{appA.3}
Since the controller cannot be efficiently expanded in the Pauli basis, we avoid performing such an expansion. Instead, we apply the techniques proposed in \cite{rahman2024feedbacke} for evaluating the controller. As discussed in Section~3 of \cite{rahman2024feedbacke}, two different approaches can be used to estimate the controller. In the first approach, the controller is expressed as:
\begin{align}    \label{uest}
\theta_{n+1} &= -\kappa \Big( \bra{\psi_n}  \mathrm{i}[H_M,H_P]  \ket{\psi_n}  \nonumber \\
& \quad \quad \quad \quad \quad \quad \quad  + 2\cdot\text{Re}\big\{ \mathrm{i}\sum_{j=0}^{m-1} \alpha_j \bra{\psi_n} H_M \ket*{z^{(j)}} \braket*{z^{(j)}}{\psi_n} \big\} \Big),
\end{align}
or in density formalism,
\begin{align}
\theta_{n+1} &= -\kappa \Big( \text{Tr}({\sigma_n}  \mathrm{i}[H_M,H_P]) \nonumber \\
&\quad \quad \quad \quad \quad \quad \quad  + 2\cdot\text{Re} \Big\{ \mathrm{i}\sum_{j=0}^{m-1} \alpha_j \text{Tr} \big( {\sigma_n} H_M \ketbra*{z^{(j)}} \big) \Big\} \Big),
\label{uestdensity}
\end{align}
where $\sigma_n=\ketbra{\psi_n}$. In this case, each term is estimated and used to compute the controller.

In the second approach, the controller is derived as the gradient of the Lyapunov function:
\begin{align}
    \label{u_dv}
     \theta_{n+1} &= - \frac{\kappa}{\Delta t} \frac{\partial}{\partial \theta_n} L_n(\theta_n),
\end{align}
where the gradient of the Lyapunov function can be approximated using finite-difference methods \cite{baydin2018automatic}:
\begin{equation}
   \frac{\partial}{\partial \theta_n} L_n(\theta_n) \approx \frac{L_n\left(\boldsymbol{\theta_n}+h \right)-L_n\left(\boldsymbol{\theta_n}-h\right)}{2h},
   \label{FDa}
\end{equation}
where $h$ is a small number. Alternatively, it can be calculated analytically using the parameter-shift rule \cite{wierichs2022general}.

\section{Control Function Design for the Feedback Law} \label{appB}
From Subsection~\ref{SS2.1}, it is seen that the function $\Lambda(\cdot)$ should be designed such that the Lyapunov function is non-increasing ($ \nabla L \cdot \dot{\sigma} \leq 0$). In Subsection~\ref{SS2.1}, the function is designed as the identity function. However, several alternative designs have been proposed in the literature. One such option is the bang-bang controller, which corresponds to optimal-time control and is defined as follows \cite{kuang2017rapid}:
\begin{equation} \label{bb}
    \Lambda_{\text{bang-bang}}(\sigma) = \text{sign} \bigg(\operatorname{Tr} \big(i\left[H_M,Q\right] \sigma\big) \bigg).
\end{equation}
It is seen that for the implementation of the bang-bang controller, it suffices to estimate the sign of the expectation. This feature is particularly advantageous when deploying these algorithms on real quantum hardware, where finite sampling introduces noise. As demonstrated in \cite{abdul2024adaptive}, this property is exploited to mitigate the sampling noise and enhance robustness, as determining the magnitude of the expectation value demands higher precision than estimating its sign. Another proposal is the finite-time controller \cite{kuang2021finite}, which is given as follows:
\begin{equation} \label{finite}
    \Lambda_{\text{finite}}(\sigma) = \text{sign} \bigg(\operatorname{Tr} \big(i\left[H_M,Q\right] \sigma\big) \bigg) \bigg| \operatorname{Tr} \big(i\left[H_M,Q\right] \sigma \bigg|^{a_1},
\end{equation}
where $a_1 \in (0,1)$. In addition, the fixed-time controller is given as \cite{li2022lyapunov}: 
\begin{align} \label{fixed}
    \Lambda_{\text{fixed}}(\sigma) &= -\kappa_1 \text{sign} \bigg(\operatorname{Tr} \big(i\left[H_M,Q\right] \sigma\big) \bigg) \bigg| \operatorname{Tr} \big(i\left[H_M,Q\right] \sigma \bigg|^{a_1} \nonumber \\
    &\;\;\;\ -\kappa_2 \text{sign} \bigg(\operatorname{Tr} \big(i\left[H_M,Q\right] \sigma\big) \bigg) \bigg| \operatorname{Tr} \big(i\left[H_M,Q\right] \sigma \bigg|^{a_2},
\end{align}
where $a_2=1/a_1$. For more details on the comparison between these design options, see \cite{abdul2024feedback}.
\section{Hotelling’s Deflation Approach} \label{appC}
Various deflation techniques, including Hotelling’s deflation, projection deflation, Schur complement deflation, and orthogonalized deflation, can be employed to handle IC constraints \cite{mackey2008deflation}. In this work, we choose Hotelling’s deflation technique since it results in a simpler quantum algorithm implementation. 

\textbf{Proposition:}  
Let $s_1 \geq s_2 \geq \dots \geq s_p$ be the eigenvalues of a Hermitian operator $S$, with corresponding normalized eigenvectors $\{\ket{s_1}, \ket{s_2}, \dots, \ket*{s_p}\}$. Define the modified operator $\bar{S} = S + \sum_{q \in \mathcal{T}} w_q \ketbra*{s_q}$, where $\mathcal{T} \subseteq \{1, 2, \dots, p\}$ is a subset of indices corresponding to the eigenvalues that are intended to be shifted, and $\{w_q\}_{q \in \mathcal{T}} > 0$ are respective shifts. Then, $\bar{S}$ retains the same eigenvectors as $S$, with eigenvalues modified as follows: 
\begin{equation}
\bar s_q(\bar{S}) = 
\begin{cases} 
s_q + w_q, & \text{if } q \in \mathcal{T}, \\
s_q , & \text{if } q \notin \mathcal{T}.
\end{cases}
\end{equation}

  \bibliographystyle{elsarticle-num} 
  \bibliography{example}







\end{document}